\documentclass[showpacs,amsmath,amssymb,prd,nofootinbib]{revtex4}
 
\usepackage{graphicx}
\usepackage{dcolumn}
\usepackage{bm}
\usepackage{epsf}
\usepackage{hyperref}
\usepackage{amsfonts}

\def\B{{\cal B}}
 
\begin{document}
 
\title{Towards a wave--extraction method for numerical 
relativity: III.~Analytical examples for the Beetle--Burko radiation scalar}

\author{Lior M.~Burko$^{1,2}$, Thomas W.~Baumgarte$^{3,4}$, and
Christopher Beetle$^{5,6}$}

\affiliation{
$^1$ Department of Physics and
Astronomy, Bates College, Lewiston, Maine 04240 \\
$^2$ Department of Physics, University of Alabama in Huntsville, 
Huntsville, Alabama 35899 \\
$^3$ Department of Physics and Astronomy, Bowdoin College,
Brunswick, Maine 04011 \\
$^4$ Department of Physics, University of Illinois at
Urbana--Champaign, Urbana, Illinois 61801 \\
$^5$ Department of Physics, Florida Atlantic University, Boca
Raton, Florida 33431 \\
$^6$ Perimeter Institute for Theoretical Physics, Waterloo, Ontario N2L 2Y5, Canada} 

\date{{\rm draft of} \today}

\begin{abstract}
Beetle and Burko recently introduced a background--independent scalar
curvature invariant for general relativity that carries information
about the gravitational radiation in generic spacetimes, in cases
where such radiation is incontrovertibly defined. In this paper we
adopt a formalism that only uses spatial data as they are used in
numerical relativity and compute the Beetle--Burko radiation scalar
for a number of analytical examples, specifically linearized
Einstein--Rosen cylindrical waves, linearized quadrupole waves, the
Kerr spacetime, Bowen--York initial data, and the Kasner spacetime.
These examples illustrate how the Beetle--Burko radiation scalar can
be used to examine the gravitational wave content of numerically
generated spacetimes, and how it may provide a useful diagnostic for
initial data sets.
\end{abstract}

 
\maketitle
 
\section{Introduction}

Compact binary systems of co-orbiting neutron stars or black holes are
among the most promising sources of gravitational waves for a set of
recently-constructed interferometric detectors, including the Laser
Interferometer Gravitational Wave Observatory (LIGO).  A substantial
effort has been underway for some years now to model these systems
theoretically, and thereby establish detailed predictions for
comparison with experimental data.  These predictions will likely be
needed not only for verification of the theory, but for constructing
filters needed to separate signal from noise in the detectors.
Because a compact binary system admits no obvious symmetry or
approximation, the theoretical problem is being tackled primarily
within the field of numerical relativity, which aims to integrate
directly the full Einstein field equations on the computer.

A numerical solution of the Einstein equations usually proceeds in three 
steps (see, e.g., \cite{bs03} for a recent review and references).
The first step solves the constraint equations of general relativity
to find initial data for the problem.  The second then evolves these
data forward in time using the remaining field equations.  In the third 
step the numerical solution of the second step is interpreted physically. 
This paper---as the other papers in this series---is interested in the 
third, last step of numerical relativity.  Although
conceptually straightforward, this program encounters many technical
difficulties which make it seem unlikely that numerical relativity
will be able to evolve compact binaries for many orbits (see, e.g.,
\cite{stu03,mdsb04,btj04} and \cite{p05} for recent progress).  Within
the field, the consensus answer to these difficulties is to start the
simulations using initial data which describe the system at small
binary separations, shortly before coalescence.  This approach
introduces a fresh conceptual problem, however.  It is far from clear
how to pick initial data at small separation so as to replicate the
outgoing wave pattern which would arise during a gradual inspiral from
large separation.  Intuitively, generic data describing a close binary
will contain spurious radiation originating in the particular
mathematical technique used to solve the constraint equations rather
than in the astrophysics of the problem.  In fact, it is known that
different methods of solving the constraint equations lead to
physically distinct initial data (see, e.g., \cite{c00,pct02,bs03}).

This paper is the third in a series which outlines a scheme to analyze
the radiation content of a given spacetime using invariant techniques
based solely on the physical metric.  More specifically, this paper
uses an invariant \textsl{radiation scalar} to measure the amount of
radiation, including spurious radiation, in a variety of initial data
sets familiar to the numerical relativity community.  In general, a
radiation scalar is a scalar function on spacetime derived from the
physical metric which, when spacetime unambiguously describes
gravitational radiation propagating in some background, manifestly
contains information only about the radiative degrees of freedom.
There may be many such metric invariants which could be defined.  We
will focus on just one, denoted $\xi$, which we call the BB radiation
scalar \cite{bb02}.  This particular scalar seems well-suited to
problems, like binary black-hole inspiral, where the quiescent
end-state describes weak radiation propagating outside a single black
hole approaching equilibrium.  Ultimately, the idea would be to
minimize in some sense this measure of radiation content and thereby
identify more physically reasonable binary initial data sets.  While
this intuitive idea is easy to state, there are a number of caveats to
taking the picture it suggests too seriously.  Nonetheless, the
results of this paper suggest the radiation scalar approach could used
effectively in this way.

The definition of the BB scalar, and the scheme on which it is based,
has certain strengths and limitations.  In this context, the key
strength of this approach lies in its direct applicability to initial
data sets. (In addition, the BB scalar may also have 
intriguing applications for evolution data.)   
The only currently-known unambiguous way to minimize the
spurious radiation content of such data is to use the evolution
equations themselves.  As evolution proceeds, spurious radiation will
naturally propagate outward, and eventually off the numerical domain
for the problem.  The gravitational radiation remaining in the
instantaneous data at later times would then be the astrophysically
sound radiation generated by the binary at earlier times.  However,
the dynamical evolution also introduces numerical error, and the
difficulties associated with this dynamical evolution, especially for
binary black hole systems, motivated the construction of
small-separation initial data in the first place.  In contrast, the BB
scalar may be calculated directly from an initial data set (see the
first paper in this series \cite{bbbn04}, hereafter Paper I).  No
evolution is required.  The approach also offers a potentially useful
tool for the extraction of gravitational radiation from numerically
evolved spacetimes (see \cite{nbbbp04}, hereafter Paper II).
However, here we focus exclusively on the application to initial data
sets.

The main risk of the BB scalar approach lies in trying to push its
physical interpretation beyond its appropriate scope.  This point has
been emphasized in previous papers \cite{bb02, bbbn04, nbbbp04} in
this series.  While mainly conceptual in origin, this issue raises
genuine practical problems in applications like those considered in
this paper.

The interpretation of the BB scalar is unambiguous when spacetime
describes weak-wave perturbations of a single black hole, an
approximation which becomes increasingly accurate at late times after
coalescence of a binary system.  In particular, it is certain to be
small in this perturbative context.  However, it remains calculable
much more generally, although this precise physical interpretation is
lost.  This is hardly surprising since even the notion of
gravitational radiation cannot be defined unambiguously in general
regions of spacetime.  More pragmatically, there is nothing to
guarantee that the radiation scalar cannot be small in this broader
context, where spacetime might intuitively be said to describe strong
radiation fields.  That is, although the radiation scalar is
definitely small when little radiation is present, it may also be
small coincidentally even when this is not the case.  This can happen
because, whereas ingoing and outgoing gravitational radiation in
perturbation theory is intuitively associated with the two
Newman--Penrose Weyl curvature components $\psi_0$ and $\psi_4$
measured in a particular null-tetrad frame, the BB scalar measures
just their product: $\xi = \psi_0\, \psi_4$.  Thus, in an extreme
example, even though the outgoing component of the radiation field
described by $\psi_4$ may be non-perturbatively large, if the ingoing
component $\psi_0$ vanishes then so does $\xi$.  However, this cannot
happen accidentally.  Intuitively, outgoing radiation will scatter off
the fixed potential described by the background in which it propagates
to produce an ingoing signal.  The outgoing signal would have to be
very carefully tuned to the background in order to have the the net
effects of this back-scattering cancel exactly throughout some region
of spacetime.  The mathematical expression of this fact lies in the
Teukolsky--Starobinsky identities, which interrelate the derivatives
of the two seemingly independent components of the radiation field
$\psi_0$ and $\psi_4$ in a Kerr black-hole background.  Following this
extreme example, one might reasonably expect situations wherein the BB
scalar is coincidentally small are rare and non-generic.  This paper
tests that expectation using exact, analytical initial data sets whose
gravitational wave content is already---independently---rather
well understood.  The results are encouraging.  We find that, at least
in these specific cases, the expectation that the radiation scalar
should not be small coincidentally in typical situations of practical
interest is indeed supported by the facts.

This paper is organized as follows.  In Section \ref{recipe}, we
summarize the recipe developed in Paper I for extracting the BB scalar
from 3+1 Cauchy data for the gravitational field.  The following three
sections explicitly calculate the BB scalar in three different
continuous families of analytical initial data sets.  Section
\ref{lin_wave} focuses on linearized wave propagating on a flat
background.  This is a worthwhile test because the BB scalar was
explicitly developed as a tool to analyze such linearized waves
propagating around a Kerr black hole.  In fact, the presence of a
dominant background component, associated with the ``Coulombic'' Kerr
geometry, has been used explicitly to define the BB scalar.  It is
therefore noteworthy that the approach is also effective when such a
dominant Coulombic field is absent.  Section \ref{rbh} turns to
genuine black hole solutions, analyzing Bowen--York initial data for
single black holes using the BB scalar.  This is probably the most
physically significant application considered here.  It finds that,
whereas the BB scalar does vanish in the Kerr spacetime where clearly
no radiation is present, it does not vanish for the Bowen--York data.
We explicitly find a quartic growth of $\xi$ with the angular momentum
$L$ of the Bowen--York hole.  This is exactly what one would naively
have expected.  Section \ref{kasner} calculates the BB scalar in the
cosmological Kasner solution.  We do this to illustrate both the
limitations and the ancillary benefits of our approach.  The BB scalar
does \textit{not} vanish in these spacetimes, even though they are
understood to contain no radiation.  However, this is not surprising
since these cosmological solutions do not contain a radiation zone in
which the usual physical meaning could be attached to the BB scalar.
Nonetheless, we find that it \textit{can} be used as an interesting
gauge-invariant tool to partially characterize the gravitational field
in this context.  One must only resist the temptation to think of it
as a \textit{radiation} scalar.  Finally, Section \ref{discsum}
summarizes our results and makes a few further comments regarding
their interpretation.

\section{Recipe}
\label{recipe}

This Section closely follows the method of paper I to construct the BB
scalar $\xi$ from numerical relativity data.  We refer to paper I
\cite{bbbn04} for further details and discussion of this recipe.  It
is recapitulated here only for convenience.

We begin our discussion at the level of spacetime geometry, where the
Weyl curvature tensor picks out exactly three null tetrads, up to
certain scaling transformations, at a generic event\relax
\footnote{Where spacetime is of algebraic Petrov types I or D, there
are three such frames.  Where spacetime is type II, there is only
one.  In the remaining cases, types III and N, no such frames exist.
These last cases, though, are far from generic, and ought not arise in
the situations of interest here.  By far, the most probable algebraic
type at a typical point of the spacetimes we study is type I.}\relax
.  These are the ``transverse frames'' in which the Newman--Penrose
curvature components $\psi_1$ and $\psi_3$ vanish.  At sufficiently
large distances from the sources, in a radiation zone, it is possible
to single out one of these three, called the quasi-Kinnersley frame,
on which our definitions are based.  The name derives from a
particular null tetrad introduced by Kinnersley on a particular
spacetime, the stationary Kerr black hole, and subsequently used by
Teukolsky to analyze the gravitational radiation content of perturbed
black hole spacetimes.  Teukolsky found that, when one makes
perturbative corrections to the background Kinnersley tetrad to
maintain vanishing Newman--Penrose curvature components $\psi_1$ and
$\psi_3$ in the perturbed spacetime, the gravitational radiation is
entirely encoded in the two curvature components $\psi_0$ and $\psi_4$
(see also \cite{szekeres}).  The remaining non-zero component,
$\psi_2$, is intuitively associated with a Coulombic part of the
field, and is dominated by terms deriving from the stationary
background.

Once the quasi-Kinnersley frame has been identified in a radiation
zone, its definition may then be carried into more general regions of
spacetime using a continuity principle.  The quasi-Kinnersley frame
is always one of the three transverse frames at \textit{any} point of
spacetime, whether that point lies in a radiation zone or not.  In
these transverse frames, the Newman--Penrose scalar $\psi_2$ will
typically take three distinct values, only one of which will lead to a
\textit{continuous} function when compared with \textit{known}
quasi-Kinnersley values of $\psi_2$ at nearby points.  The frame so
chosen is the quasi-Kinnersley frame.  Put another way, the key idea
is that there is no ambiguity in identifying the quasi-Kinnersley
frame in a radiation zone, and maintaining continuity everywhere in
spacetime generally determines a unique ``transverse frame field''
throughout.  This is the quasi-Kinnersley frame, defined globally.
The results of this paper will indicate that there ought to be little
doubt in practice concerning which value of $\psi_2$ will give the
desired continuous extension.  See the Discussion for further
comments.

This paper will consider two curvature scalars defined using the
quasi-Kinnersley frame:
\begin{equation}\label{csDefs}
	\chi := \psi_2 \qquad\mbox{and}\qquad \xi := \psi_0\, \psi_4.
\end{equation}
Each of these functions is understood to be defined on spacetime by
evaluating the indicated Newman--Penrose curvature components
\textit{in the quasi-Kinnersley frame}.  The first of these
quantities, $\chi$, is called the Coulomb scalar.  The nomenclature
derives from intuition in the Teukolsky formalism: $\psi_2$ is
dominated by the stationary, Coulombic field of the background Kerr
black hole.  The second quantity, $\xi$, is the BB radiation scalar.
In a radiation zone accurately described by the Teukolsky formalism,
it manifestly depends only on the radiative degrees of freedom.
Although the components $\psi_0$ and $\psi_4$ are not separately
preserved by the residual scaling transformations of the null tetrad
allowed by the Weyl curvature, their product is.  Thus, the BB scalar
is a true invariant derived from spacetime curvature.

Although certainly not a radiation scalar, the Coulomb scalar is still
quite useful in our approach.  First, as described above, demanding
continuity of $\chi$ plays the key role in defining the
quasi-Kinnersley frame.  We could also have used continuity of $\xi$
to do this, but $\xi$ is quadratic in the curvature while whereas
$\chi$ is only linear.  The radiation scalar therefore falls of much
more rapidly ($\xi \sim r^{-6}$) at large distance than the Coulomb
scalar ($\chi \sim r^{-3}$), making the numerical problem of
determining continuity much harder.  Second, the Coulomb scalar can be
used to extract information about the single black hole underlying a
perturbation-theoretic description of a given spacetime.  Thus, for
example, it may be used at late times to determine, or at least
constrain estimates of, the mass and spin of the final black hole
resulting from coalescence in a background-independent way.

Numerical relativity is not primarily concerned with spacetime
geometry.  Rather, it solves a Cauchy problem for general relativity,
calculating a series of spatial geometries parameterized by a fiducial
time variable.  These may then be assembled to form spacetime.  The
assembly procedure can of course be done analytically, and used to
express the BB scalar in terms of given Cauchy data.  This has been
done in Paper I, and we describe the results here.

In a $3+1$ decomposition, the gravitational fields are expressed in
terms of a spatial metric $\gamma_{ij}$ and an extrinsic curvature
$K_{ij}$.  The electric and magnetic components $E_{ij}$ and $B_{ij}$
of the spacetime Weyl tensor can be computed from these data using
\begin{eqnarray} 
{\cal E}_{ij} & = & R_{ij} + K K_{ij} - K_{ik} K_{j}^{~k} - 4 \pi S_{ij} \\
E_{ij} & = & {\cal E}_{ij} - \frac{1}{3} \gamma_{ij} \gamma^{kl} {\cal E}_{kl}
	\label{E}
\end{eqnarray}
where $S_{ij} = \gamma_i^{~a} \gamma_{j}^{~b} T_{ab}$ is the spatial
projection of the stress energy tensor $T_{ab}$, and 
$R_{ij}$ is the spatial Ricci tensor (see, e.g., Ref.~\cite{bs03} for 
more detail). The definition for the 
magnetic part of the spacetime Weyl tensor depends on the convention 
used for the extrinsic curvature and for the spacetime volume element. 
We adopt here the convention that 
\begin{equation}\label{K-def}
K_{ij}=\,-\,\frac{1}{2}{\cal L}_{n}\,\gamma_{ij}\, ,
\end{equation}
where ${\cal L}_n$ denotes the Lie derivative in the direction of the 
normal $n^{a}$ to the 3--hypersurface. Our convention for the 
spacetime volume element is
\begin{equation}
\epsilon_{abcd}=\, 24\,i\,\ell_{[a}\,n_{b}\,m_{c}\,{\bar m}_{d]}
\end{equation}
(see paper I for more detail.) We next define the magnetic part of the 
spacetime Weyl tensor
\begin{eqnarray} 
	{\cal B}_{ij} & = & - \epsilon_i^{~kl} \nabla_k K_{lj }\\
	B_{ij} & = & {\cal B}_{(ij)}.  \label{B}
\end{eqnarray}
By construction, $E_{ij}$ and $B_{ij}$ are both symmetric and
traceless.  We then define the complex 
tensor\footnote{Note that the 
relative sign in the definition of $C^i{\,_j}$ here 
differs from that used in Paper I.  At first glance, it would appear that  
the quantities we calculate here are therefore the complex conjugates  
of those calculated there.  The difference, however, is illusory.   
Our previous papers have tacitly assumed a   
definition for the extrinsic curvature with the opposite sign to that  
in Eq.~(\ref{K-def}).  While the previous convention is popular in the  
mathematical relativity community (see, e.g., \cite{wald}), the  
current one is used more often in the numerical relativity  
community.  We therefore adopt it here for the convenience of the latter.  
As a result of this shift of convention,  
the sign of the tensor ${\cal B}_{ij}$ defined above is reversed.  We  
compensate this reversal in the definition of $C^i{}_j$ so that this  
tensor takes the exact same value on a Cauchy slice with our new  
conventions as it did with the old.  With this one modification, all  
other formulae from the previous papers continue to hold unmodified.}
\begin{equation}\label{C}
	C^i_{~j} = E^i_{~j} \,+\, i\, B^i_{~j},
\end{equation}
from which the scalar curvature invariants $I$ and $J$ can be computed as 
\begin{equation} \label{I}
	I = \frac{1}{2} C^i_{~j}  C^j_{~i}
\end{equation}
and
\begin{equation} \label{J}
	J = - \frac{1}{6}  C^i_{~j} C^j_{~l} C^l_{~i}.
\end{equation}
Once a basis is chosen, the tensor $C^i_{~j}$ becomes simply a $3
\times 3$ matrix.  In ordinary linear algebra, it is well-known that
any scalar invariant which may be extracted from such a matrix must be
expressible as a function of three basic invariants.  These basic
invariants may be chosen to be the eigenvalues or, for example, the
traces of the first three powers of the matrix in question.  The trace
of our matrix vanishes because it derives from the spacetime Weyl
tensor, and the traces of its square and cube are basically $I$ and
$J$.  Thus, we should expect our Coulomb and BB scalars may be
expressed as functions of $I$ and $J$.  To do this, we recall the
Baker--Campanelli speciality index \cite{bc00} defined by
\begin{equation} \label{S}
	S := 27 \frac{J^2}{I^3}.
\end{equation}
In terms of $S$, the Coulomb scalar can be written as
\begin{equation} \label{chi}
\chi^{0,\pm} = - \frac{3 J}{2 I} \frac{W_{\chi}(S)^{1/3} + 
W_{\chi}(S)^{-1/3}}{\sqrt{S}},
\end{equation}
where $W_{\chi}(S) = \sqrt{S} - \sqrt{S - 1}$, while the BB scalar is
\begin{equation} \label{xi}
\xi^{0,\pm} = \frac{1}{4}\, I\, \left[2-W_{\xi}(S)^{1/3}-
W_{\xi}(S)^{-1/3}\right]\, ,
\end{equation}
where $W_{\xi}(S)=2S-1+2\sqrt{S(S-1)}$.  In general, both the Coulomb
and BB scalars admit three distinct complex roots, associated with the
three distinct transverse frames.  We denote these three roots with
the superscripts ``$0,\pm$''.  The value in the principal branch of
the root functions above has superscript $0$, and this branch defines
the quasi-Kinnersley frame in an appropriate radiation zone.  The
other two branches have superscripts $\pm$.  Either of these may be
labeled the quasi-Kinnersley value in the interior of spacetime,
depending on what is needed to maintain continuity while coming in
from the radiation zone.  We will illustrate this point explicitly
below.

In addition to the direct calculation above, there are two other ways
to compute $\xi^{0, \pm}$ and $\chi^{0, \pm}$.  The first is to solve
the eigenvalue problem for the complex spatial tensor $C^i_{~j}$:
\begin{equation}\label{evProb}
	C^i_{~j}\, {\hat \sigma}^j = \lambda\, {\hat \sigma}^i = \left( 2\chi^{0, \pm} \right) \hat\sigma^i.
\end{equation}
As shown in Paper I, the three eigenvalues $\lambda$ turn out to be
exactly twice the Weyl scalar $\psi_2$ evaluated in the three
transverse frames.  Thus, the Coulomb scalar at any point of
spacetime is always half of one of the eigenvalues of $C^i_{~j}$.  If
these eigenvalues are calculated, this approach does not say which is
which, so it is not immediately clear which eigenvalue is associated
with the principal branch of (\ref{chi}), and therefore with the
quasi-Kinnersley frame.  However, it has been shown in Paper II that,
among the three branches, the principal value has the largest modulus
in a radiation zone.  Thus, the prescription to calculate the Coulomb
scalar is to take half the eigenvalue of largest modulus at large
distances, and extend inward to strong-field regions always choosing
the unique eigenvalue which keeps the function $\chi$ smooth.  In a
transverse frame, the Newman--Penrose Weyl scalars $\psi_1$ and
$\psi_3$ vanish by definition, and this gives a particularly simple
relation between the Coulomb and BB scalars:
\begin{equation}\label{alternative_xi}
\xi^{0,\pm} = (\,\psi_0\,\psi_4\,)^{0,\pm} = 
I - 3\left(\psi_2^{0,\pm}\right)^2.
\end{equation}
Thus, if the Coulomb scalar is known, the radiation scalar follows
immediately.  The second way to calculate the Coulomb and BB scalars
is to use series expansions of the exact formulae above.  As shown in
paper I, for a spacetime that asymptotically is perturbed about
algebraic speciality, we can expand
\begin{equation}\label{chi_expansion}
\chi =-3\,\frac{J}{I}\, \left[1-\frac{4}{9}\,(S-1)+
\frac{80}{243}\,(S-1)^2+\cdots\,\right],
\end{equation}
and
\begin{equation}\label{xi_expansion}
\xi =-\frac{I}{9}\,(S-1)\,\left[1-\frac{8}{27}\,(S-1) + 
\frac{112}{729} \,(S - 1)^{2}\cdots\,\right].
\end{equation}
These expansions explicitly use the principal values of the root
functions in (\ref{chi}) and (\ref{xi}), and therefore allow us to
identify the quasi-Kinnersley $\chi$ and $\xi$ asymptotically .  This
technique can be useful numerically when all three values $\chi^{0,
\pm}$ become small.  We illustrate this fact in Sections
\ref{bowen-york} for Bowen-York data and
\ref{kasner} for the Kasner spacetime.

\section{Linearized waves on a flat background}
\label{lin_wave}

Probably the first example that comes to mind when evaluating a
radiation scalar are pure wave solutions.  Unfortunately, the simplest
such solution, plane waves, is too simple to be of interest.  Plane
waves are Petrov type N, and hence algebraically special.  In a type N
spacetime, a frame can always be found so that $\psi_4$ is the only
non-vanishing Weyl scalar.  This frame is automatically transverse, so
that $\xi = \psi_0 \psi_4$ vanishes identically.  As the BB scalar is
frame-independent, it vanishes also in all other, non-canonical
frames.  In a plane wave spacetime the BB scalar vanishes despite the
presence of gravitational radiation.

We therefore turn to two less trivial examples, namely linearized
time-symmetric Einstein--Rosen cylindrical gravitational waves in
Section \ref{einstein-rosen} and linearized even-parity quadrupole
Teukolsky waves in Section \ref{teukolsky}.

\subsection{Linearized Einstein--Rosen cylindrical waves}
\label{einstein-rosen}

The metric for Einstein--Rosen cylindrical waves of amplitude $B$ and
pulse width $a$ \cite{weber57} is given by
\begin{equation}\label{E-R_metric}
ds^2 = \,e^{2\,[\gamma(t,\rho)-\psi(t,\rho)]}
\left(-\,dt^2\,+\,d\rho^2\,\right) 
+\,e^{2\psi(t,\rho)}\,dz^2+
\,\rho^2\,e^{-2\psi(t,\rho)}\,d\phi^2 \, ,
\end{equation}
where 
\begin{equation}
\psi(t,\rho) = B\,
\left[\,\frac{1}{\sqrt{(a+it)^2+\rho^2}}+\,
\frac{1}{\sqrt{(a-it)^2+\rho^2}}\,\right]\,
\end{equation}
(which is a real function.)  In what follows we consider linearized
waves, so that we can take the function $\gamma(t,\rho)$, which is
quadratic in $B$, to vanish.  The solution is time-symmetric about
$t=0$, and describes cylindrical waves that are imploding on the
symmetry axis for $t<0$ and exploding for $t>0$.

In the context of numerical relativity, the spacetime metric 
is usually expressed in terms of a lapse $\alpha$, a shift $\beta^i$ 
and a spatial metric $\gamma_{ij}$.  These quantities can be found
be identifying the spacetime metric (\ref{E-R_metric}) with the 
``ADM'' form 
\begin{equation}\label{adm}
ds^2 = - \alpha^2 dt^2 + \gamma_{ij} (dx^i + \beta^i dt)(dx^j + \beta^j dt)\, ,
\end{equation}
Here the shift vanishes identically, $\beta^i = 0$, the lapse is
\begin{equation}
\alpha = \exp [-\psi(t,\rho)]\,,
\end{equation}
and the spatial metric is
\begin{equation}
\,d\sigma^2 = \,e^{2\,[-
\psi(t,\rho)]}\,d\rho^2 +\,e^{2\psi(t,\rho)}\,dz^2+
\,\rho^2\,e^{-2\psi(t,\rho)}\,d\phi^2 \,.
\end{equation}
We also compute the extrinsic curvature $K_{ij}$ from Eq.~(\ref{K-def}) 
\begin{equation} \label{ext_curv}
K_{ij}= - \frac{1}{2\alpha} \left( \partial_{t}\gamma_{ij}
- D_i \beta_j - D_j \beta_i \right),
\end{equation}
where $D_i$ is the covariant derivative compatible with the spatial
metric $\gamma_{ij}$, and where we use the sign convention typically
used in numerical relativity.

\begin{figure}
\input epsf
\centerline{ \epsfxsize 10.0cm \epsfbox{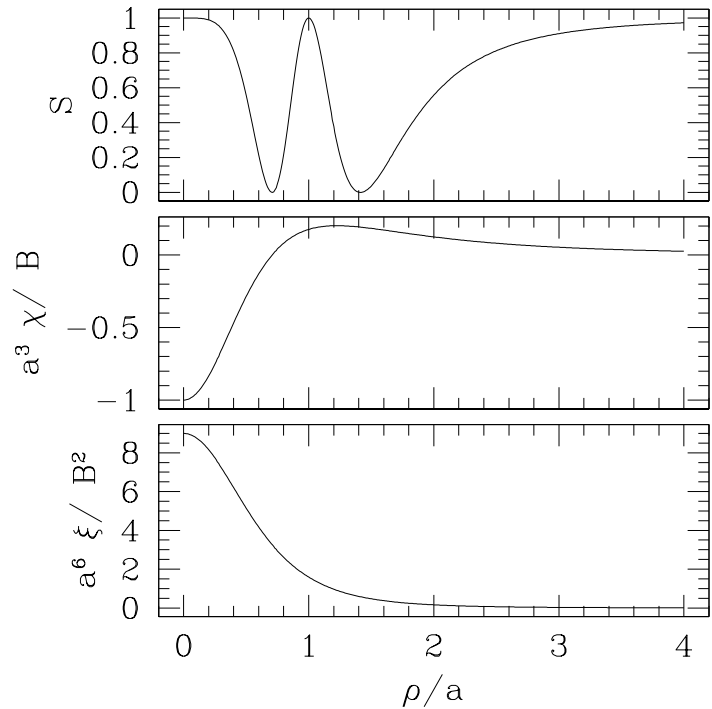}}
\caption{The speciality index $S$ (upper panel), the dimensionless
Coulomb scalar $a^3\,\chi/B$ (middle panel), and the dimensionless BB
scalar $a^6\,\xi/\,B^2$ (lower panel), as functions of $\rho/a$ for
time-symmetric linearized Einstein--Rosen cylindrical waves of pulse
width $a$, at $t=0$.  }
\label{fig:ER}
\end{figure}

At the moment of time symmetry $t=0$, $K_{ij}$ vanishes identically,
which greatly simplifies the problem.  The only non-vanishing term in
the electric (\ref{E}) or magnetic (\ref{B}) components of the Weyl
tensor is then the Ricci tensor $R_{ij}$ in the electric part $E_{ij}$,
which yields
\begin{eqnarray}
E^\rho_{\;\rho} &=& -2B\,\left(\,\rho^2+a^2\,\right)^{-3/2}\nonumber \\
E^z_{\;z} &=& 2B\,\left(2a^2-\rho^2\right)\,
\left(\,\rho^2+a^2\,\right)^{-5/2}\nonumber \\
E^{\phi}_{\;\phi} &=& 2B\,\left(2\rho^2-a^2\right)\,
\left(\,\rho^2+a^2\,\right)^{-5/2}
\end{eqnarray} 
to linear order in $B$ (all other components vanish.)  At $t=0$ we
therefore have $C^i_{~j}\,=\,E^i_{~j}$.  Using (\ref{I}) and (\ref{J})
we find the curvature invariants
\begin{equation}
I = 12B^2\,\frac{a^4-a^2\rho^2+\rho^4}{(\rho^2+a^2)^5}
\end{equation}
and
\begin{equation}
J = -4B^3\,\frac{(2\rho^2-a^2)\,(\rho^2-2a^2)}{(\rho^2+a^2)^{13/2}}\,,
\end{equation}
and from (\ref{S}) the specialty index
\begin{equation}\label{speciality_ER}
S=\frac{(2\rho^2-a^2)^2\,(\rho^2-2a^2)^2\,
(\rho^2+a^2)^2}{4(a^4-a^2\rho^2+\rho^4)^3}\, .
\end{equation}
We plot the specialty index $S$ as a function of $\rho$ in the upper
panel of Fig.~\ref{fig:ER}.  Interestingly, $S$ is independent of the
wave amplitude $B$ and therefore does not provide a measure of the
strength of the gravitational radiation.

Since $S$ is independent of $B$, the Coulomb scalar $\chi$ (\ref{chi})
must scale with $B$ and the BB scalar $\xi$ (\ref{xi}) with $B^2$.
Reversing this argument we can see that the speciality index $S$ must
be independent of the wave amplitude whenever the Coulomb scalar
is perturbative and scales with $B$, while the BB scalar scales
with $B^2$.  We find the same behavior for the Teukolsky waves in
Section \ref{teukolsky}.

For the remainder of this Section we determine the Coulomb scalar
$\chi$ and the BB radiation scalar $\xi$ explicitly as functions of
$\rho$, adopting the following strategy.  There are three regions for
which the spacetime is close to speciality, i.e., for which $S - 1 \ll
1$: $\rho\gg a$, $\rho=a$ and $\rho\ll a$ (compare Fig.~\ref{fig:ER}.)
We then identify $\chi$ with an eigenvalue of
$C^i_{~j}$. Specifically, we require that at asymptotically great
distances ($\rho\gg a$) the expansion (\ref{chi_expansion}) holds, and
then identify $\chi$ with an eigenvalue of $C^i_{~j}$ --- which is
equivalent to picking a particular branch in the cubic roots in
(\ref{chi}).  From $\chi$ we can then compute $\xi$ using
(\ref{alternative_xi}).  Finally, we require continuity and
differentiability as we move to smaller and smaller values of $\rho/a$.

Before we proceed we list the three eigenvalues of $C^i_j$, which we denote
$\lambda_\rho,\,\lambda_z$ and $\lambda_{\phi}$,
\begin{eqnarray}\label{eigenvalues}
\lambda_\rho &=& -2B\,\left(\,\rho^2+a^2\,\right)^{-3/2}\nonumber \\
\lambda_z &=& 2B\,\left(2a^2-\rho^2\right)\,
\left(\,\rho^2+a^2\,\right)^{-5/2}\nonumber \\
\lambda_{\phi} &=& 2B\,\left(2\rho^2-a^2\right)\,
\left(\,\rho^2+a^2\,\right)^{-5/2}\, .
\end{eqnarray}
Notice that $I=(1/2)\sum_i\lambda_i^2$ and
$J=(1/2)\Pi_i\lambda_i$ as expected.

To find $\chi$ and $\xi$ we examine the asymptotic region
$\rho\gg a$. In this limit,
\begin{equation}
I \, \approx \, \frac{12}{\rho^6}\,B^2\, 
~~~~~~~~~~~~~~~~\rho\gg a, 
\end{equation}
and 
\begin{equation}
J \, \approx \, -\frac{8}{\rho^9}\,B^3 
~~~~~~~~~~~~~~~~\rho\gg a, 
\end{equation}
so that (\ref{chi_expansion}) yields  
\begin{equation}
\chi\,\approx\, -3\frac{J}{I}\, \approx\, \frac{2}{\rho^3}\,B\, .
~~~~~~~~~~~~~~~~\rho\gg a, 
\end{equation}
This the quasi-Kinnersley $\chi$. We compare this result with the
limiting values of the three eigenvalues (\ref{eigenvalues}) of the
tensor $C^i_{\;j}$,
\begin{equation}
\lambda_\rho  \,\approx\, -\frac{2}{\rho^3}\, B,~~~~
\lambda_z     \,\approx\, -\frac{2}{\rho^3}\, B,~~~~
\lambda_{\phi}\,\approx\,  \frac{4}{\rho^3}\,B\,
~~~~~~~~~~~~~~~~\rho\gg a,
\end{equation}
The eigenvalue that corresponds to $\chi$ is the one that satisfies
$\lambda = 2 \chi$, and therefore we conclude that for $\rho\gg a$,
the correct eigenvalue is $\lambda_{\phi}$.  (Alternatively, this is
the eigenvalue with the greatest modulus --- see paper II.)  Next, we
require that $\chi$ is smooth to extend $\chi$ to other regions of the
spacetime. In principle, we could switch to a different eigenvalue at a
branch point.  This would make $\chi$ discontinuous, however, which is
not desirable.  A distributional $\chi$ would suggest similar
characteristics of the physical Weyl scalar, which is clearly not the
case in this example.  The only identification for $\xi$ that makes it
smooth throughout is that $\chi=\lambda_{\phi}/2$ not just for
$\rho\gg a$, but everywhere.  We therefore have
\begin{equation}
\chi \, = \,\frac{2\rho^2-a^2}{(\rho^2+a^2)^{5/2}}\, B\, 
\end{equation}
everywhere.  We can now compute the BB radiation scalar $\xi$ from
(\ref{alternative_xi}), which yields
\begin{equation}
\xi \,=\,I-3\,\chi ^2\,=\,9\,\frac{a^4}{(\rho^2+a^2)^5}\, B^2 \, .
\end{equation}
Our results for $\chi$ and $\xi$ are plotted in Fig.~\ref{fig:ER} in
the middle and lower panels, correspondingly.

In Appendix \ref{ER} we present an alternative, yet equivalent,
derivation of $\chi$ and $\xi$, that is based directly on
Eqs.~(\ref{chi}) and (\ref{xi}), respectively.

\subsection{Linearized quadrupole waves}
\label{teukolsky}

Einstein--Rosen cylindrical waves are not asymptotically flat, which
limits their usefulness as testbeds for numerical relativity.  (In
addition, Einstein--Rosen cylindrical waves do not represent the
exterior solution of bounded radiating systems.)  We therefore briefly
discuss a second example of linearized gravitational waves over a flat
background, namely Teukolsky's even-parity linearized quadrupole
gravitational waves in vacuum with azimuthal symmetry
\cite{teukolsky-82}.  This spacetime is algebraically-general (Petrov
type I), that asymptotically at great distances ($\lambda^{1/2}\,r\gg
1$) approaches algebraic speciality (Petrov type III). This algebraic
class will be important for the interpretation of the speciality index
$S$ and the BB scalar $\xi$.

From the spacetime metric (Eq.~(5) in \cite{teukolsky-82}) we find the
lapse $\alpha$ to be unity, the shift $\beta^i$ to vanish, and we can
identify the spatial metric $\gamma_{ij}$ as
\begin{eqnarray}\label{teukolsky_metric}
\,d\sigma^2&=&
[1+(2-3\,\sin^2\theta)\,{\cal A}]\,dr^2
-6\,{\cal B}\,r\,\sin^2\theta\,\cos\theta\,dr\,d\theta
+(1-{\cal A}+3\,\sin^2\theta\,{\cal C})\,r^2\,d\theta^2
\nonumber \\
&+&[1+(3\sin^2\theta-1)\,{\cal A}
-3\,\sin^2\theta\,{\cal C}]\,r^2\,\sin^2\theta\,d\phi^2.
\end{eqnarray}
Here the coefficients ${\cal A}$, ${\cal B}$ and ${\cal C}$ are given
in terms of a function $F(x) := F(t \mp r)$ and its derivatives
$F^{(n)} \equiv \,d^nF(x)/\,dx^n |_{x = t \mp r}$,
\begin{eqnarray}\label{teukolsky_k}
{\cal A}_{\mp}&=&3\left(\frac{F^{(2)}}{r^3} \pm \frac{3 F^{(1)}}{r^4}
+ \frac{3 F}{r^5}\right)\\
{\cal B}_{\mp}&=& -\left( 
\pm \frac{F^{(3)}}{r^2}+ \frac{3F^{(2)}}{r^3} \pm
\frac{6 F^{(1)}}{r^4}+ \frac{6 F}{r^5}\right)\\
{\cal C}_{\mp}&=& \frac{1}{4}\,\left( \frac{F^{(4)}}{r} \pm
\frac{2 F^{(3)}}{r^2}+\frac{9F^{(2)}}{r^3} \pm
\frac{21 F^{(1)}}{r^4}+ \frac{21F}{r^5}\right),
\end{eqnarray}
where the upper/lower signs correspond to that in the definition of
$F$.  We would like to construct time-symmetric data with $K_{ij} = 0$
at $t=0$, and therefore choose linear combinations ${\cal A}={\cal
A}_--{\cal A}_+$, ${\cal B}={\cal B}_--{\cal B}_+$, and ${\cal
C}={\cal C}_--{\cal C}_+$. For concreteness, we choose the function
\begin{equation}
F(x) = \epsilon \, (t \pm r)^5 \, e^{- \lambda (t \pm r)^2}
\end{equation}
where $\epsilon$ is a measure of the wave amplitude and $\lambda$ is
related to the wavelength.

As for Einstein--Rosen waves, the vanishing of the extrinsic curvature
at $t=0$ greatly simplifies the problem.  Since the spatial Ricci
tensor $R_{ij}$ is linear in $\epsilon$, we have ${\cal
E}_{ij}=R_{ij}$ to leading order in $\epsilon$ and also $B_{ij}=0$.
This scaling is sufficient to determine the scaling of the BB
radiation scalar $\xi$: Since $C^i_{~j} = E^i_{~j}$ is linear in
$\epsilon$, $I$ must scale with $\epsilon^2$ and $J$ with
$\epsilon^3$.  We again find that the speciality index $S$ is
independent of $\epsilon$.  It does depend on the choice of $F(x)$, but
is independent of the wave amplitude.  That is, $S$ is insensitive to
the amplitude of the waves. As we demonstrate in Appendix
\ref{app:quad}, the speciality index $S\to 0$ as $\lambda^{1/2}\,
r\to\infty$. However, as spacetime is asymptotically neither Petrov
types 0 nor N, while $I,J\to 0$, it is algebraically special and of
Petrov type III.  This property is not accounted for by the speciality
index $S$, which does not approach unity as $\lambda^{1/2}\,
r\to\infty$.

The Coulomb scalar $\chi$ again scales with the wave amplitude
parameter $\epsilon$, while the BB scalar $\xi$ scales with its
square, $\epsilon^2$.

Explicit expressions for the Coulomb and BB radiation scalar can be
constructed with an analysis similar to that in Section
\ref{einstein-rosen} for Einstein--Rosen waves.  The calculations are
significantly more involved, however, and do not necessarily provide
new insight.  Instead of going through these calculations we assure
the reader that the results are qualitatively very similar to those
for Einstein--Rosen waves, and refer to Appendix \ref{app:quad} for
some details.

\section{Rotating Black Holes}
\label{rbh}

\subsection{Kerr black holes}
\label{kerr}

The Kerr solution is of Petrov type D and hence algebraically 
special with
\begin{equation}
S = 1,
\end{equation}
independently of the black hole angular momentum $L = aM$.  From
Eq.~(\ref{xi_expansion}) it is evident that the BB radiation scalar
vanishes, $\xi = 0$.  This can also be seen by observing that
$W_{\xi}=1$ when $S=1$.  The three cubic roots of unity are
$1$,$e^{2i\pi/3}$ and $e^{4i\pi/3}$, which, when inserted into
$\xi^{0,\,\pm}$, yield zero and twice $3I/4$.  These are the three
roots found in paper II, and the BB radiation scalar corresponds to
the vanishing value.  Given that the Kerr spacetime is stationary and
does not contain any gravitational radiation it is reassuring that the
radiation scalar vanishes.

Even though we already know the result, it is useful to demonstrate
that the recipe laid out in Section \ref{recipe} would lead to the
same conclusion.  To do so we start with the spacetime metric of a
Kerr black hole, which in Boyer-Lindquist coordinates can be written
as
\begin{equation}
ds^2 = - \frac{\rho^2 \Delta}{\Sigma}\, dt^2 + \frac{\Sigma}{\rho^2}
\sin^2 \theta \left( d \phi - \frac{2 M a R}{\Sigma} dt \right)^2
+\frac{\rho^2}{\Delta} \, dR^2 + \rho^2 \, d \theta^2
\end{equation}
where we have used the abbreviations $\Delta=R^2-2MR+a^2$,
$\rho^2=R^2+a^2\,\cos^2\theta$, and
$\Sigma=(R^2+a^2)(R^2+a^2\,\cos^2\theta)+2a^2MR\,\sin^2\theta$.  As in
Section \ref{einstein-rosen} we can identify the spatial metric of
constant Boyer--Linquist time slices as
\begin{equation} \label{kerr-metric}
\gamma_{ij} dx^i dx^j = \frac{\rho^2}{\Delta} dR^2 + \rho^2 d\theta^2 +
\frac{\Sigma}{\rho^2} \sin^2\theta d\phi^2,
\end{equation}
the lapse as
\begin{equation}
\alpha=\sqrt{\frac{\rho^2\Delta}{\Sigma}}
\end{equation}
and the shift vector as
\begin{equation}
\beta_{a}=(0,0,-2MaR\rho^{-2}\sin^2\theta).
\end{equation}
From (\ref{ext_curv}) we then find the only non-vanishing components
of the extrinsic curvature (up to symmetry) to be
\begin{equation} \label{kerr_K_rphi}
K_{R\phi}=\frac{aM\{3R^2\rho^2+
a^2[\rho^2-2(R^2+a^2)\,\cos^2\theta]\}\,\sin^2\theta}
{\rho^3\Delta^{1/2}\Sigma^{1/2}}
\end{equation}
and
\begin{equation} \label{kerr_K_thetaphi}
K_{\theta\phi}=-2\frac{ 
Ma^3R\Delta^{1/2}\,\sin^3\theta\,\cos\theta}{\rho^3\Sigma^{1/2}}\, .
\end{equation}

The electric part of the Weyl tensor now can be computed from (\ref{E})
\begin{eqnarray}
E_{RR}&=&-\frac{MR(2\Sigma +3a^2\Delta\,\sin^2\theta ) (\rho^2 
-4a^2\,\cos^2\theta)}{\Delta\rho^4\Sigma}\\
E_{R\theta}&=&3\frac{a^2M(R^2+a^2)(3\rho^2-4a^2\,\cos^2\theta 
)\,\sin\theta\,\cos\theta}{\rho^4\Sigma}\\
E_{\theta\theta}&=&\frac{MR(\rho^2 
-4a^2\,\cos^2\theta)(\Sigma+3a^2\Delta\,\sin^2\theta)}{\rho^4\Sigma}\\
E_{\phi\phi}&=&\frac{MR\Sigma (\rho^2 -4a^2\,\cos^2\theta )\,\sin^2\theta }
{\rho^8}
\end{eqnarray}
and the magnetic part from (\ref{B})
\begin{eqnarray}
B_{RR}&=&-\frac{aM(3\rho^2 -4a^2\,\cos^2\theta)(2\Sigma+3a^2\Delta 
\,\sin^2\theta )}{\Delta\rho^4\Sigma}\,\cos\theta \\
B_{R\theta}&=&-3\frac{aMR(R^2+a^2)(\rho^2 
-4a^2\,\cos^2\theta )}{\rho^4\Sigma}\,\sin\theta \\
B_{\theta\theta}&=&-\frac{aM(3\rho^2 -4a^2\,\cos^2\theta 
)(\Sigma +3a^2\Delta\,\sin^2\theta)}{\rho^4\Sigma}\,\cos\theta \\
B_{\phi\phi}&=& \frac{aM\Sigma (3\rho^2 -4a^2\,\cos^2\theta 
)}{\rho^8}\,\sin^2\theta\,\cos\theta\, .
\end{eqnarray}
From $E_{ij}$ and $B_{ij}$ we construct $C_{ij}$ as well as the
invariants $I$ and $J$, following Section \ref{recipe}.  As expected
we find
\begin{equation}
S=1\, ,
\end{equation}
which is by no means surprising but reassuring.

\subsection{Bowen--York Initial Data}
\label{bowen-york}

Initial Data describing rotating black holes can also be constructed
by solving the constraint equations in the Bowen--York \cite{by80}
formalism.  This approach assumes that the spatial metric is
conformally flat.  For rotating Kerr black holes, slices of constant
Boyer--Linquist time are not conformally flat, nor are axisymmetric
foliations that smoothly reduce to slices of constant Schwarzschild
time in the Schwarzschild limit \cite{gp00}.  This suggests that
Bowen--York initial data give rise to a spacetime that is distinct from
the Kerr-Schild spacetime.  This was demonstrated explicitly by
Gleiser, Nicasio, Price and Pullin (\cite{gnpp98}, hereafter GNPP),
who used a perturbative calculation to show that Bowen--York initial
data evolve into a spacetime that can be interpreted as a Kerr black
hole with gravitational radiation.  GNPP found that the amplitude of
the emitted gravitational radiation scales with the square of the
black hole's angular momentum $L$, and the power accordingly with
$L^4$.  As an important test we verify in this Section that the BB 
scalar picks up this gravitational radiation and identifies the 
correct scaling.

In a conformal transverse-traceless decomposition of the vacuum
constraint equations the Hamiltonian constraint reduces to
\begin{equation} \label{ham1}
8 \bar \nabla^2 \psi - \psi {\bar R} - \frac{2}{3} \psi^5 K^2 + \psi^{-7}
\bar A_{ij} \bar A^{ij} = 0
\end{equation}
and the momentum constraint is
\begin{equation} \label{mom1}
\bar \nabla_j \bar A^{ij} - \frac{2}{3} \psi^6 \bar \gamma^{ij} \bar
\nabla_j K = 0
\end{equation}
(see, e.g., \cite{c00,bs03,p04} for recent reviews).  The conformal
factor $\psi$ relates the physical spatial metric $\gamma_{ij}$ to the
conformally related metric $\bar \gamma_{ij}$ via
\begin{equation}
\gamma_{ij} = \psi^4 \bar \gamma_{ij},
\end{equation}
where $\bar \nabla_i$ and $\bar R$ are the covariant derivative and scalar
curvature associated with $\bar \gamma_{ij}$, and the extrinsic
curvature $K_{ij}$ is decomposed into its trace $K$ and a traceless
part $\bar A_{ij}$ as
\begin{equation} \label{K_decomp}
K_{ij} = \psi^{-2} \bar A_{ij} + \frac{1}{3} \gamma_{ij} K.
\end{equation}

Assuming conformal flatness ($\bar \gamma_{ij} = \eta_{ij}$, where
$\eta_{ij}$ is the flat spatial metric in any coordinate system) and
maximal slicing results in $\bar R = 0$ and $K = 0$, so that equations
(\ref{ham1}) and (\ref{mom1}) reduce to
\begin{equation} \label{ham2}
\bar \nabla^2 \psi = - \frac{1}{8} \psi^{-7} \bar A_{ij} \bar A^{ij}
\end{equation}
and
\begin{equation} \label{mom2}
\bar \nabla_j \bar A^{ij} = 0.
\end{equation}
Here $\bar \nabla_i$ is now the flat covariant derivative associated
with $\eta_{ij}$ (as we assumed ${\bar \gamma}_{ij}=\eta_{ij}$).

The momentum constraint (\ref{mom2}) decouples from the Hamiltonian
constraint (\ref{ham2}) under these assumptions and can be solved
analytically.  For a spinning, unboosted black hole in polar
coordinates (i.e. $\bar \gamma_{ij} = \mbox{diag}(1,r^2,
r^2\sin^2\theta)$) the only non-vanishing component of $\bar A_{ij}$
is
\begin{equation} \label{ang_mom}
\bar A_{r\phi} = \frac{3 \sin^2 \theta}{r^2} L
\end{equation}
where $L$ is the angular momentum \cite{by80}.  

Given the angular momentum $L$ and Eq.~(\ref{ang_mom}), the conformal factor
$\psi$ can be found from Eq.~(\ref{ham2}).  Since the angular momentum $L$
only enters squared in the Hamiltonian constraint (\ref{ham2}), the
conformal factor $\psi$ is even in $L$.  In general, a solution to this
quasi-linear elliptic equation has to be constructed numerically, but
an approximate solution up to $O(L^4)$ is given by
\begin{equation} \label{psi}
\psi = 1 + \frac{M}{2r} + \left(\frac{L}{M^2}\right)^2 
\left(1 + \frac{M}{2r} \right)^{-5} 
\left[ \tilde\psi_0 + \tilde\psi_2 P_2(\cos \theta) \right] + F(r,\theta)\,L^4
\end{equation}
where
\begin{equation}
\tilde\psi_0 =
- \frac{M}{5 r}
    \left[ 5 \left(\frac{M}{2r}\right)^3 
    + 4 \left(\frac{M}{2r}\right)^4 
    + \left(\frac{M}{2r}\right)^5 \right]
\end{equation}
and
\begin{equation}
\tilde\psi_2 = - \frac{1}{10} \left(\frac{M}{r}\right)^3 
\end{equation}
and where the $P_2(\cos \theta) = (3 \cos^2 \theta - 1)/2$ is the
second Legendre polynomial (see GNPP.)  We allow for an {\em unknown} 
$O(L^4)$ contribution by including the unknown function $F(r,\theta)$
that, in principle, can be found by solving the Hamiltonian constraint
to $O(L^4)$ in a method similar to that of GNPP.  

The extrinsic curvature, which must be odd in $L$, can then be found from 
Eq.~(\ref{K_decomp}) 
\begin{eqnarray} 
K_{r\phi} & = & \psi^{-2} \bar A_{r\phi} \nonumber \\
& = & 12 \frac{\,\sin^2\theta}{(2 r + M)^2}L 
- \frac{48}{5}\,
\frac{[8(1-3\,\cos^2\theta)r^3-20Mr^2-8M^2r-M^3]\,\sin^2\theta}
{(2r + M)^8} L^3 + O(L^5)\, . \label{BY_K}
\end{eqnarray}

Even though we will find the BB radiation scalar $\xi$ is of order
$O(L^4)$, it turns out that it depends only on terms in $\psi$ up to
order $O(L^2)$, so that it is independent of the unknown function
$F(r,\theta)$.  A heuristic argument for this behavior can be given as
follows (for a more rigorous proof see Appendix \ref{app:BY_proof}.)
The Bowen--York solution is equivalent to the Kerr solution up to
$O(L)$, and deviations enter only at order $O(L^2)$.  In a transverse
frame we therefore expect the leading-order terms in the Weyl scalars
$\psi_0$ and $\psi_4$, which are measures of the gravitational
radiation that is absent in the Kerr solution, to be $O(L^2)$.  The
leading order term in the radiation scalar $\xi$, being the product of
$\psi_0$ and $\psi_4$) should therefore be $O(L^4)$, and higher order
corrections in the conformal factor should enter at higher order.

\begin{figure}
\input epsf
\centerline{ \epsfxsize 10.0cm
\epsfbox{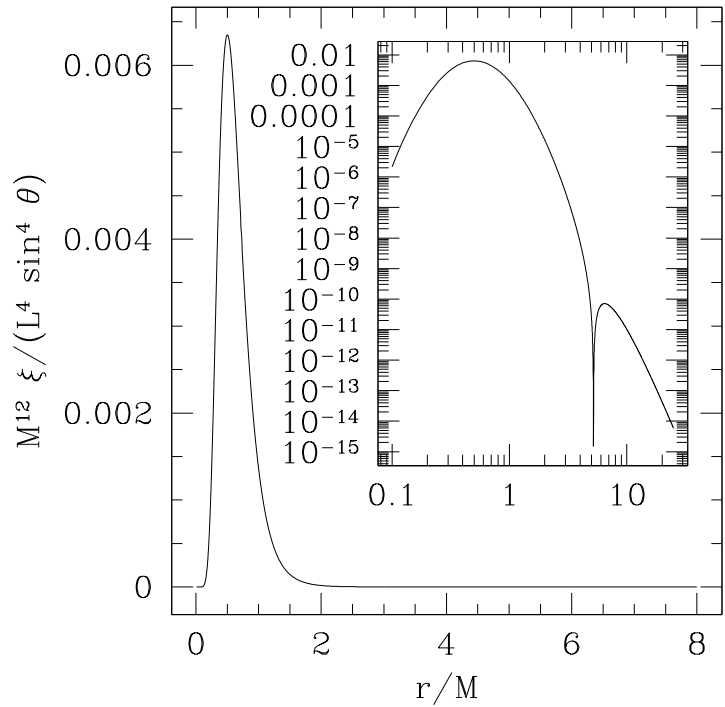}}
\caption{The dimensionless BB scalar $M^{12}\xi/(L^4\,\sin^4\theta)$
for Bowen--York initial data as a function of $r/M$. (For
$\theta=0,\pi$, the BB scalar $\xi=0$.)  }
\label{fig:BY}
\end{figure}

To construct $\xi$, we first compute the electric (\ref{E}) and
magnetic (\ref{B}) parts of the Weyl tensor.  We list the results in
Appendix \ref{app:B-Y}, and point out that these quantities do depend
on $F(r,\theta)$.  We then compute the matrix $C^i_{~j}$ and, using
(\ref{I}) and (\ref{J}), the invariants $I$ (\ref{I_expansion}) and
$J$ (\ref{J_expansion}).  Again, $I,J$ depend on $F(r,\theta)$.  From
(\ref{S}) we find the speciality index $S$ to $O(L^4)$ 
\begin{equation}
S = 1  - \frac{3^3\,  
2^{10}}{5^2}\, \frac{(4r^2-21Mr+M^2)^2r^4\,\sin^4\theta}
{M^4(2r+M)^{12}}\, L^4\, ,
\end{equation}
which turns out to be independent of $F(r,\theta)$.  Because we were
able to express the speciality index as a deviation from unity---i.e.,
we expressed spacetime as a deviation from algebraic speciality---we
can find the BB scalar $\xi$ immediately from (\ref{xi_expansion})
to $O(L^4)$
\begin{equation}
\xi = \frac{3^2 \, 
2^{22}}{5^2}\, \frac{(4r^2-21Mr+M^2)^2r^{10}\,\sin^4\theta}
{M^2 (2r+M)^{24}}\, L^4 
\end{equation}
(see Fig.~\ref{fig:BY}). To $O(L^2)$ the Coulomb scalar $\chi$ is
given by
\begin{eqnarray}
\chi  = &-& 64\,\frac{Mr^3}{(2r+M)^6}-768\,i\,
\cos\theta\frac{r^4}{(2r+M)^8}\, L 
+ \frac{768}{5}\,(24r^4+8r^4\,\cos^2\theta+24r^3M\,\cos^2\theta\nonumber \\
&-&8Mr^3+2r^2M^2\,\cos^2\theta
-14M^2r^2-8M^3r-M^4)\,\frac{r^3}{M(2r+M)^{12}}\, L^2\, .
\end{eqnarray}

As expected, the BB radiation scalar scales with $L^4$.  Since the
deviation between Bowen--York data and a Kerr spacetime enters at
order $L^2$, the amplitude of the emitted gravitational radiation also
scales with $L^2$, and the power carried by the gravitational waves
accordingly with $L^4$ (see GNPP.)  As we have seen for the linearized
waves on flat backgrounds in Section \ref{lin_wave}, the BB radiation
scalar scales with the square of the gravitational wave amplitude.  At
great distances ($r\gg M$), the BB scalar $\xi\sim L^4/(M^2\,r^{10})$.
The peeling off of the BB scalar with $r^{-10}$ is understood as
follows: The BB scalar is expected to drop like the product of
$\psi_0$ and $\psi_4$.  The usual peeling off of the product is
expected to scale like $M^2/r^6$.  However, in the Kerr spacetime at
that order it would be multiplied by a {\em zero} coefficient.  The
next order, which would peel off like $ L^2/r^8$, also vanishes
because the deviation of the Bowen--York initial data from Kerr is at
second-order in $L$ in the metric functions.  The next term is indeed
the one found.  This scaling is quadratic in $\chi-\chi_{\rm Kerr}$,
as expected.  When $\xi$ is calculated by finding the difference
between $I$ and $\chi^2$, the Kerr contributions to $\chi$ cancel with
the Kerr contributions to $I$, so that the BB scalar $\xi$ depends
only on the radiative Bowen--York degrees of freedom.

Notice, that $\xi$ peaks at $r=M/2$, i.e., at the Schwarzschild black
hole's event horizon.  The radiation described by the BB scalar inside
the event horizon evidently cannot change the exterior
spacetime. However, because the greatest deviations of the Bowen--York
initial data from Kerr are localized near the event horizon, it is
expected that much of this spurious radiation be absorbed by the black
hole.

\section{Kasner spacetime}
\label{kasner}

The Kasner spacetime is a homogeneous vacuum spacetime.  From its
spacetime metric we identify a unity lapse, a vanishing shift vector,
and the spatial metric
\begin{equation} \label{kasner-metric}
\,d\sigma^2=t^{2p_1}\,dx^2+t^{2p_2}\,dy^2+t^{2p_3}\,dz^2\, ,
\end{equation}
where the parameters $p_{i}$ satisfy
$p_1+p_2+p_3=1=p_1^2+p_2^2+p_3^2$.  It is convenient express $p_2$ and
$p_3$ in terms of $p_1$, which we will call $p$, so that
\begin{equation} 
p \equiv p_1~~~~~~
p_{2,3}=(1-p\pm\varrho)/2,
\end{equation}
where $\varrho=[(1-p)(1+3p)]^{1/2}$.  The Kasner spacetime is
sufficiently simple to construct the BB radiation scalar in closed
form, and also to demonstrate the construction of the quasi-Kinnersley
frame from spatial data as they are used in numerical relativity.

We compute the extrinsic curvature from the spatial metric and find
\begin{equation} \label{kasner-k}
K_{a b} = {\rm diag}[\,-p\,t^{2p-1}\,,\,-(1-p+\varrho)\,t^{-p+\varrho}/2\,,\,
-(1-p-\varrho)\,t^{-p-\varrho}/2\,]\, ,
\end{equation}
so that $K=-1/t$.  For the Kasner spacetime the magnetic part of the
Weyl tensor vanishes, so that $C_{a b}=E_{a b}$.  The only non-vanishing
components are
\begin{eqnarray}
C^x_{~x} &=& p\,(1-p)\,t^{-2} \nonumber \\
C^y_{~y} &=& (1-p+\varrho)\,(1+p-\varrho)\,t^{-2}/4 \\
C^z_{~z} &=& (1-p-\varrho)\,(1+p+\varrho)\,t^{-2}/4 \nonumber \, ,
\end{eqnarray}
From (\ref{I}) and (\ref{J}) we now find the Weyl tensor invariants
\begin{eqnarray}
I &=& \frac{p^2(1-p)}{t^4} \\
J &=& \frac{p^4(1-p)^2}{2t^6}\, ,
\end{eqnarray}
and from (\ref{S}) the speciality index
\begin{equation} \label{S_kasner}
S=\frac{27}{4}(1-p)p^2\, .
\end{equation}

Since the matrix $C^a_{~b}$ is diagonal, its three eigenvalues and
corresponding eigenvectors can be found quite easily.  In order to
find which one of these corresponds to the quasi-Kinnersley frame we
parametrize $p$ as $p=-1/3+q$ where $q$ does not need to be small.
Inserting this into (\ref{S_kasner}) we find
\begin{equation}
S - 1 = - \frac{27}{4} \,q + \frac{54}{4} \,q^2
- \frac{27}{4} \,q^3.
\end{equation}
The case $q=0$ corresponds to an algebraically special
spacetime (Petrov type D), and for small values of $q$ we 
can therefore use the expansion (\ref{chi_expansion}) to find 
\begin{equation}
\chi =-\frac{2}{9\,t^2}\,\left(1-\frac{15}{4}\, q
+\frac{9}{4}\, q^2+\cdots\right)
\end{equation}
For small values of $q$ this coincides with $\lambda_x/2$, 
so that we can identify
\begin{equation}
\chi=-\frac{p_2\, p_3}{2\,t^2}\, .
\end{equation}
We can find the BB scalar $\xi$ either from the series expansion
(\ref{xi_expansion}), or, having found $\chi$ already, from
(\ref{alternative_xi}), which yields
\begin{equation}
\xi =\frac{q}{9t^4}\,(1-3\,q )^2
\left(1-\frac{3}{4}\,q\right)
= \frac{1}{4\,t^4}\,p_1^2\,(p_2-p_3)^2\, .
\end{equation}
Evidently we have $\xi\ne 0$, even though no radiation is present in
the Kasner spacetime.  [There are three cases for which $\xi=0$: $q=0$
(spacetime is Petrov type D), and $q=1/3$ or $4/3$, which are flat
spacetimes (Petrov type 0).]  While this seems very counter-intuitive,
we remind the reader that the BB radiation scalar is an unambiguous
measure of gravitational radiation {\em only} in a radiation zone,
which the homogeneous Kasner solution does not possess.  The
{\it name} ``radiation scalar" is therefore may be misleading in this case. 

Our results for $I,J$, and $\chi$ for the Kasner spacetime agree with
those obtained through a pure spacetime approach by Cherubini {\em et
al.}  \cite{cherubini}, who also compute the Weyl scalars $\psi_0$ and
$\psi_4$ in a transverse frame, and their product agrees with our
value of $\xi$, as expected.

We now demonstrate how the quasi-Kinnersley frame can be constructed
from spatial data.  The eigenvector that corresponds to the eigenvalue
$\lambda_x$, and hence to the quasi-Kinnersley frame, is
\begin{equation}
{\hat \sigma}^a=s\,\delta^a_x\, ,
\end{equation}
where $s = t^{-p_1}$ is a normalization constant. Evidently ${\hat
\sigma}^a$ is real and has vanishing imaginary part.

Following paper I, the two real projections onto the hypersurface of
the two real null vectors of the spacetime tetrad are
\begin{equation}
\hat{\lambda}^a=t^{-p_1}\,\delta^a_x\;\;\;\;\;\;\;\;\;\;
\hat{\nu}^a=-t^{-p_1}\,\delta^a_x\, ,
\end{equation}
so that, with the unit normal to the hypersurface
$\hat{\tau}^a=\delta^a_t\,\partial_a$, the two real spacetime null
vectors of the frame are, up to a spin--boost parameter $c$, given by
\begin{eqnarray}
\ell^a &=& \frac{|c|}{\sqrt{2}}(\,\partial_t+t^{-p_1}\,\partial_x)
\label{kasner_l} \\
n^a &=& \frac{|c|^{-1}}{\sqrt{2}}(\,\partial_t-t^{-p_1}\,\partial_x)\, . 
\label{kasner_n}
\end{eqnarray}
Notice, that the two real projections of $\ell^a,n^a$ on the
hypersurface are anti-parallel. This happens because by construction the
normal $\hat{\tau}^a$ lies in the spacetime tangent 2-plane spanned
by $\ell^a$ and $n^a$. In this degenerate case, the complex basis
vector $m^a$ can be found as follows (see paper I for details). We
first choose an arbitrary real unit vector ${\hat r}^a$ in the spatial
2-plane orthogonal to ${\hat\lambda}^a=-{\hat\nu}^a$. Then,
\begin{equation}
m^a=\frac{e^{i\vartheta}}{\sqrt{2}}\, \left({\hat
r}^a+i\,\varepsilon^{abc}\,{\hat \lambda}_b\,{\hat r}_c\,\right)\, .
\end{equation}
We find that $\varepsilon^{abc}=-[a\;b\;c]/\sqrt{{\rm
det}(g_{mn})}=-[a\;b\;c]/t$, with $[a\;b\;c]$ being the permutation
symbol. We next choose ${\hat r}^a=t^{-p_2}\,\delta^a_y$, so that
\begin{equation}\label{kasner_m}
m^a=\frac{e^{i\vartheta}}{\sqrt{2}}\,\left(\,t^{-p_2}\,\partial_y-\,i\,t^{-p_3}
\,\partial_z\,\right)\, .
\end{equation}
The vectors (\ref{kasner_l}),(\ref{kasner_n}), and (\ref{kasner_m})
are the vectors that make the quasi-Kinnersley frame.  Note that we
are able to determine the quasi-Kinnersley frame only up to spin-boost
with parameter $c=|c|\exp(i\vartheta)$. In some situations one can 
choose the spin-boost parameter based on the physical properties of 
spacetime, so that the quasi-Kinnersley tetrad can be 
found (see, e.g., Ref.~\cite{berti05}). 
However, this problem awaits further study.

\section{Summary and Discussion}
\label{discsum}

The BB radiation scalar $\xi$ was introduced as an invariant measure
of gravitational radiation in regions of spacetime where such
radiation is unambiguously defined.  However, it is uniquely and
smoothly extended throughout a generic spacetime.  In this paper, we
have computed $\xi$ for a variety of analytical spacetimes and initial
data sets, adopting a formalism that relies only on spatial data as
they are typically used in numerical relativity.  These calculations
have illustrated the procedure used to define $\xi$ explicitly.  That
procedure has been described previously \cite{bb02, nbbbp04, bbbn04},
but not implemented.  Its actual implementation here, therefore,
should help clarify the procedure in general.

For those examples we have considered here which unambiguously contain
gravitational radiation, namely linearized waves on flat backgrounds
and Bowen--York initial data for rotating black holes, our radiation
scalar scales with the square of parameters that govern the
gravitational wave amplitude.  This suggests that it does indeed
provide a reasonable measure of the gravitational wave content of
these data.  (The speciality index $S$, in contrast, is independent of
the waves' amplitude when the Coulombic part of curvature is
perturbative.)  We also have examined the homogeneous, cosmological
Kasner solutions.  In this case we found a non-vanishing $\xi$ even
though Kasner does not contain gravitational radiation.  We explain
this apparent contradiction by noting the absence of a radiation zone
in the Kasner spacetime.  Thus, although the {\it name} ``radiation
scalar'' for $\xi$ is misleading and inappropriate in this case, $\xi$
can still be found uniquely.  From the mathematical point of view, it
remains an interesting gauge-invariant ``observable'' even when it is
not a \textit{radiation} scalar.

This paper has illustrated the potential of $\xi$ as a tool for
discriminating among initial data sets for numerical relativity.
However, it has done so using analytic, rather than actual numerical,
data.  An actual application to such numerical data sets would
engender a fresh set of difficulties, originating in the fundamental
discreteness of the variables involved and the associated subtlety in
obtaining the required \textit{smooth} extension of the
quasi-Kinnersley frame into the strong-field regions of the initial
data set.  A full discussion of these issues lies outside the scope of
this paper, and will be addressed elsewhere.  However, we must make
some comment here.  To illustrate the resolution we envision to these
difficulties in actual numerical work, we examine the results of
Figure \ref{fig:ER-3chi} in the Appendix.  These figures show all
three values of the Coulomb and BB scalars, appropriately scaled to
give dimensionless quantities, for linearized Einstein--Rosen waves of
flat spacetime.  At large radii, one branch of the upper graph, the
dotted one, is clearly larger than the other two.  This is the
principal branch of $\xi$, and is associated with the quasi-Kinnersley
frame in this region.  Moving inward, there are three critical points
where we must be careful, at $\rho/a = \sqrt{2}$, $1$ and
$1/\sqrt{2}$.

The outermost critical point poses no real problem since the root we
had been tracking, $\xi^0$, is visibly discontinuous there.  If we
were moving inward in a numerical spacetime, calculating eigenvalues
of the tensor $C^i_{~j}$ with sufficiently small radial steps, there
is no way we would be confused about which eigenvalue was appropriate
just inside $\rho/a = \sqrt{2}$.  Inside this first critical point, we
would naturally be tracking the dashed root, rather than the dotted.

The next critical point, at $\rho/a = 1$ poses a greater difficulty.
The discontinuity there occurs in the derivative of the Coulomb
scalar, rather than in the scalar itself, and could therefore be
harder to detect in numerical data.  However, even in this case, there
is a simple resolution to the problem.  We would be able to recognize
numerically that the eigenvalue we are interested in, associated with
the dashed curve, is degenerating with another, associated with the
solid curve.  There are then at least two things we could do to make
sure the Coulomb scalar remains smooth at this critical point.  One
would be to use higher-order approximants, such as splines based on
$\chi$-values at several grid points in each direction around the
critical point, to explore which branch we ought to pick to preserve
smoothness.  An even simpler approach would be to exploit the known
analytic structure of the Riemann surface underlying the multi-valued
complex function underlying $\chi$.  That structure is described in
detail in Papers I and II.  Essentially, this approach would boil down
to calculating how the speciality index $S$ varies in a neighborhood
of the critical point, and mapping spacetime in that neighborhood
into the Riemann surface.  The critical point itself occurs where $S$
lies exactly on the branch line we have chosen for $\chi$.  As $S$
crosses that branch line, there is no ambiguity at this level in how
the branch of $\chi$ ought to change to preserve analyticity.  Thus,
once we have detected the possible presence of such a critical point
in numerical data, we could use the explicit formula of (\ref{chi})
and the local structure of the function $S$ on spacetime to predict
the exact eigenvalue which corresponds to the quasi-Kinnerlsey frame
at every grid-point nearby.

The third critical point, at $\rho/a = 1/\sqrt{2}$ poses no difficulty
whatsoever.  At this point, the quasi-Kinnersley value is associated
with the solid curve, and it is the other two which are discontinuous.
We would scarcely notice this in actual numerical work.

The Einstein--Rosen example discussed here is actually quite useful.
The three critical points we have observed in this example typify the
only three kinds of critical points which can occur in a general
spacetime, whether numerical or analytic in origin.  The above
discussion shows how these three types of critical point may be
handled using a combination of simple numerical tests and insights
from the analytical development of the radiation scalar approach.
Although further testing will doubtless be required in developing
general numerical implementations, these observations are encouraging
for the ultimate viability of this approach in numerical relativity.

A second question which lies outside the scope of the present paper,
but which we nonetheless ought to address briefly here, concerns how
the radiation scalar could potentially be used in discriminating among
potential choices of initial data.  Consider, for example, initial
data describing a compact binary at small binary separation.  As we
have discussed before, these data will presumably contain
``astrophysically sound'' radiation together with the ``junk
radiation'' that we would like to minimize.  Such initial data can be
constructed using different approaches, namely different
decompositions of the constraint equations and different choices for
the freely specifiable background geometry.  Different choices lead to
physically distinct solutions \cite{c00,pct02,bs03} with distinct
gravitational wave content.  Our comparison of Kerr and Bowen--York
data for rotating black holes in Section \ref{rbh} illustrates how the
BB radiation scalar may be used as a diagnostic for comparing the
amount of gravitational waves that these initial data sets would emit
when evolved dynamically.  

We find in the analytical examples considered here that the BB
radiation scalar scales as expected with a parameter that controls the
gravitational wave content of these examples.  Such a parameter may
not exist in numerically generated initial data sets, in which case it
may be less clear how any non-zero value of $\xi$ should be
interpreted.  The BB radiation scalar may nevertheless be a useful
diagnostic.  Even for a single initial data set the ratio $\xi/\chi^2$
would provide a dimensionless measure of the strength of the
``radiative'' fields compared to the ``coulombic'' fields.  Perhaps
more importantly, it may be useful to compare $\xi$ for different
initial data sets that approximate the same physical situation.  For
example, such a comparison may provide some guidance for deciding
which choices in the construction of initial data sets yields a more
truthful representation of binary black holes or neutron stars at
small binary separations.  We can envision various different ways of
how such a comparison could be made.  One potentially useful approach
would focus on the radiation zone, where we have reliable intuition
about how the real radiation content ought to fall off.  Comparing the
fall-off behavior with that of the BB radiation scalar may help
distinguish this ``astrophysically sound'' radiation content from the
``junk'' content and minimize the latter.

As a word of caution, however, we point out that such a comparison can
at best provide some guidance rather than conclusive evidence.  It is
possible that the individual radiation fields $\psi_0$ and $\psi_4$
change in such a way that their product $\xi = \psi_0 \psi_4$
decreases even though the true radiation content increases.  However,
for many generic sets (or families) of initial data this may not be the
case.  Since a diagnostic providing conclusive evidence is not yet
available, we therefore believe that the BB radiation scalar may in the
meantime provide some useful guidance as long as we keep these caveats
in mind.

We finally discuss some further limitations of this approach.  While
the BB radiation scalar $\xi$ contains some measure of gravitational
wave content, it is not clear immediately how this information can be
translated into gravitational wave templates that might be useful for
gravitational wave observers.  Specifically such observers would
presumably measure the gravitational wave amplitudes $h_\times$ and
$h_+$ in a transverse-traceless frame.  In the linearized examples we
have studied here, both the intuitive gravitational wave content and
the BB radiation scalar scales with the square of a parameter
describing intuitively the strength of the waves being modeled.  This
does not mean, however, that $\xi$ is proportional to the energy of
the gravitational wave.  In general the latter scales with $|\psi_0 +
\psi_4|^2$ in the appropriate basis, while $\xi = \psi_0 \psi_4$.
Since $\psi_0$ and $\psi_4$ have different fall-off behavior, $\xi$
also falls off differently from both the gravitational wave amplitude
or energy. In short, although $\xi$ describes only the radiative
degrees of freedom, $\xi$ does not carry with it {\em all} the
information---or even sufficient information---about the
gravitational waves required by observers.  This issue will be
discussed more fully in a forthcoming paper.

\acknowledgments

The authors are indebted to Marco Bruni and Andrea Nerozzi for
invaluable discussions.  TWB gratefully acknowledges support from the
J.~S.~Guggenheim Memorial Foundation.  This work was supported in part
by NSF Grant PHY--0139907 to Bowdoin College, by NSF grant
PHY--0400588 to Florida Atlantic University, and by NASA grant
ATP03-0001-0027 through the University of Texas at Brownsville.

\begin{appendix}

\section{Direct calculation for Einstein--Rosen waves}\label{ER}

In this appendix we find $\chi$ and $\xi$ for Einstein--Rosen
cylindrical waves directly from Eqs.~(\ref{chi}) and
(\ref{xi}). First, we re-write Eq.~(\ref{chi}) as
\begin{equation}
\chi^{0,\pm} \,=\, -\,\frac{3}{2}\,\frac{J}{I}\,Z^{0,\pm}_{\chi}(S)\,
,
\end{equation}
where
\begin{equation}\label{branches}
Z^{i}_{\chi}(S) \,=\,
\frac{1}{\sqrt{S}}\,\left[\,\alpha_i\left(\,\sqrt{S}\,-\,\sqrt{S-1}\,\right)^{1/3}\,
+\,\alpha_i^{-1}\left(\,\sqrt{S}\,-\,\sqrt{S-1}\,\right)^{-1/3}\,\right]\,
\end{equation}
Here, $\alpha$ is one of the three cubic roots of unity, $\alpha_0=1$,
$\alpha_+=\exp\, (2i\pi/3)$, and $\alpha_-=\exp\, (4i\pi/3)$. By the
cubic and square roots in Eq.~(\ref{branches}) we mean the principal
branch roots for either.  The choice of the cubic root then is done
through the choice of $\alpha_i$. This choice labels the three
different branches of the function $Z^{i}_{\chi}(S)$.  The
corresponding functions $\xi^{0,\pm}$ can be calculated using
\begin{equation}
\xi^{0,\pm}\,=\,I-3\,(\chi^{0,\pm})^2\, ,
\end{equation}
or, equivalently, from Eq.~(\ref{xi}) using a similar approach to its
three branches.

\begin{figure}
\input epsf
\centerline{ \epsfxsize 10.0cm
\epsfbox{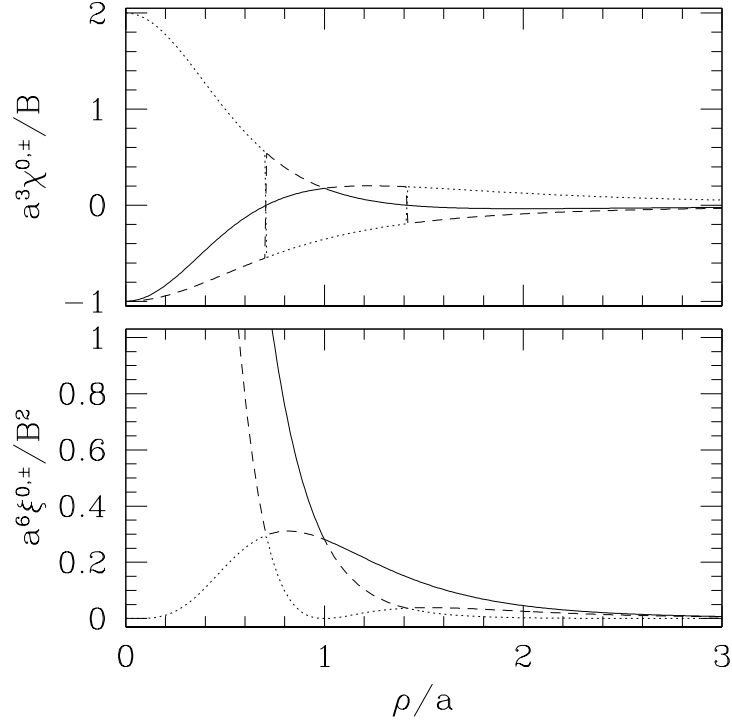}
}
\caption{The three branches of the Coulomb scalar $\chi^{0,\pm}$ (top
panel) and the BB scalar $\xi^{0,\pm}$ (bottom panel) for linearized
Einstein--Rosen cylindrical waves.  The principal branch---$\chi^0$
and $\xi^0$---are dotted, $\chi^-$ and $\psi^-$ are dashed, and
$\chi^+$ and $\psi^+$ are the solid curves.  }
\label{fig:ER-3chi}
\end{figure}

Figure \ref{fig:ER-3chi} shows the three different branches for
$\chi^{0,\pm}$ and $\xi^{0,\pm}$, correspondingly. Notice that the
three branches for either function are non-differentiable, and for
$\chi^0$ and $\chi^-$ even discontinuous. For either function,
however, branches can be changed so that the resulting functions are
smooth.  The identity of the physically meaningful functions can be
determined by demanding a proper asymptotic behavior of $\chi$ and
$\xi$.  In this case, starting at great distances, the chosen branch
is the principal branch, $\chi^0$ (and correspondingly, $\xi^0$). At
the branch point $\rho/a=\sqrt{2}$ we change the branch to $\chi^-$
(and, correspondingly, $\xi^-$), and at $\rho/a=1$ we change the
branch to $\chi^+$ (and, correspondingly, $\xi^+$). The resulting {\em
smooth} curves are identical with those presented in
Fig.~\ref{fig:ER}.

\section{Some expressions for linearized quadrupole waves}
\label{app:quad}

From the spatial Ricci tensor we find that at the moment of time 
symmetry, $t=0$, the only non-vanishing 
components of $C^i_{~j}$ are to leading order in $\epsilon$,
\begin{eqnarray}
C^r_{\;r} &=& 24\,(25\,r^{6}\,\lambda ^{3} - 2\,r^{8}\,\lambda ^{
4} + 54\,r^{2}\,\lambda  - 78\,r^{4}\,\lambda ^{2} + 224\,r^{4}\,
\lambda ^{2}\,\cos^2\theta - 73\,r^{6}\,\lambda ^{3}
\,\cos^2\theta \nonumber \\
& &- 158\,r^{2}\,\lambda \,\cos^2\theta + 6\,r
^{8}\,\lambda ^{4}\,\cos^2\theta + 4\,\cos^2\theta )\,
e^{- r^{2}\,\lambda }\,\epsilon /r^{2} \\
C^{\theta}_{\;r} & = & r^{-2}\,C^{r}_{\;\theta} = 24\,\sin\theta\,\cos\theta \,
( - 12 + 293\,r^{6}\,\lambda ^{3} - 64\,r^{8}\,\lambda ^{4} + 150\,r^{2}\,
\lambda  + 4\,r^{10}\,\lambda ^{5} - 420\,r^{4}\,\lambda ^{2})\,
e^{- r^{2}\,\lambda }\,\epsilon / r^3 \\
C^{\theta}_{\;\theta} &=&   
- 24\,(6 + 715\,r^{6}\,\lambda ^{3} - 290\,r^{8}\,
\lambda ^{4} + 78\,r^{2}\,\lambda  + 292\,r^{8}\,\lambda ^{4}\,
\cos^2\theta - 739\,r^{6}\,\lambda ^{3}\,\cos^2\theta  \nonumber \\
&& - 567\,r^{4}\,\lambda ^{2} + 640\,r^{4}\,\lambda ^{2}\,
\cos^2\theta  + 43\,r^{10}\,\lambda ^{5} - 43\,
\cos^2\theta \,r^{10}\,\lambda ^{5} - 130\,r^{2}\,
\lambda \,\cos^2\theta  \nonumber \\
&&- 4\,\cos^2\theta - 2\,r^{12}\,\lambda ^{6}
 + 2\,\cos^2\theta \,r^{12}\,\lambda ^{6})\,e^{ - r^{2
}\,\lambda }\,\epsilon /r^{2} \\
C^{\phi}_{\;\phi} &=& 
24\,(6 + 690\,r^{6}\,\lambda ^{3} - 288\,r^{8}\,
\lambda ^{4} + 24\,r^{2}\,\lambda  + 286\,r^{8}\,\lambda ^{4}\,
\cos^2\theta - 666\,r^{6}\,\lambda ^{3}\,\cos^2\theta \nonumber \\
&& - 489\,r^{4}\,\lambda ^{2} + 416\,r^{4}\,\lambda ^{2}\,
\cos^2\theta + 43\,r^{10}\,\lambda ^{5} - 43\,
\cos^2\theta \,r^{10}\,\lambda ^{5} + 28\,r^{2}\,
\lambda \,\cos^2\theta \nonumber \\
&& - 8\,\cos^2\theta - 2\,r^{12}\,\lambda ^{6}
 + 2\,\cos^2\theta \,r^{12}\,\lambda ^{6})\,e^{- r^{2}\,\lambda }\,
\epsilon /r^{2}
\, ,
\end{eqnarray}
Restricting analysis to the equatorial plane, $\theta=\pi /2$, we find
the curvature scalars
\begin{eqnarray}
I&=&
576\,(36 - 1548\,r^{4}\,\lambda ^{2} - 51744\,r^{6}
\,\lambda ^{3} + 353559\,r^{8}\,\lambda ^{4} - 27664\,r^{18}\,
\lambda ^{9} - 172\,r^{22}\,\lambda ^{11} 
- 451732\,r^{14}\,\lambda ^{7} + 612\,r^{2}\,\lambda \nonumber \\ 
&& -
773889\,r^{10}\,\lambda ^{5} + 803755\,r^{12}\,\lambda ^{6} +
3005\,r^{20}\,\lambda ^{10} 
+ 146051\,r^{16}\,\lambda ^{8} + 4\,r^{24}\,\lambda ^{12}
)\,e^{- 2\,r^{2}\,\lambda }\,\epsilon ^{2}/r^{4}
\end{eqnarray}
and
\begin{eqnarray}
J&=&
- 6912\,\lambda \,(
2\,r^{6}\,\lambda ^{3} - 54 + 78\,r^{2}\,\lambda  - 25\,r^{4}\,
\lambda ^{2}) \nonumber \\
&& \times ( - 6 - 690\,r^{6}\,\lambda ^{3} + 288\,r^{8}\,\lambda ^{4} - 24
\,r^{2}\,\lambda  - 43\,r^{10}\,\lambda ^{5} + 489\,r^{4}\,
\lambda ^{2} + 2\,r^{12}\,\lambda ^{6})\nonumber \\
&& \times ( - 6 - 715\,r^{6}\,\lambda ^{3} + 290\,r^{8}\,\lambda ^{4} - 78
\,r^{2}\,\lambda  - 43\,r^{10}\,\lambda ^{5} + 567\,r^{4}\,
\lambda ^{2} + 2\,r^{12}\,\lambda ^{6})\,e^{ - 3\,r^{2}\,\lambda }
\,\epsilon ^{3}/r^{4}\, 
\end{eqnarray}
From these we compute the speciality index
\begin{eqnarray}
S &=&
27\,r^4\,\lambda ^{2}\,(2\,r^{6}\,\lambda ^{3} - 54 +
78\,r^{2}\,\lambda  - 25\,r^{4}\,\lambda ^{2})^{2}
( - 6 - 690\,r^{6}\,\lambda ^{3} + 288\,r^{8}\,\lambda ^{4} - 24
\,r^{2}\,\lambda  - 43\,r^{10}\,\lambda ^{5} + 489\,r^{4}\,
\lambda ^{2} + 2\,r^{12}\,\lambda ^{6})^{2} \nonumber \\
&&\times  ( - 6 - 715\,r^{6}\,\lambda ^{3} + 290\,r^{8}\,\lambda ^{4} - 78
\,r^{2}\,\lambda  - 43\,r^{10}\,\lambda ^{5} + 567\,r^{4}\,
\lambda ^{2} + 2\,r^{12}\,\lambda ^{6})^{2} \nonumber \\
& \left/
{\vrule height0.50em width0em depth0.50em} \right. \!  \! &
\left[ 4( 36 - 1548\,r^{4}\,\lambda ^{2} - 51744\,r^{6}\,\lambda ^{3} +
353559\,r^{8}\,\lambda ^{4} - 27664\,r^{18}\,\lambda ^{9} - 172\,
r^{22}\,\lambda ^{11} 
- 451732\,r^{14}\,\lambda ^{7} + 612\,r^{2}\,\lambda \right. \nonumber \\
& & -  \left. 
 773889\,r^{10}\,\lambda ^{5} + 803755\,r^{12}\,\lambda ^{6} +
3005\,r^{20}\,\lambda ^{10} + 146051\,r^{16}\,\lambda ^{8} 
+ 4\,r^{24}\,\lambda ^{12})^{3} \right] \, .
\end{eqnarray}
As expected, $S$ is independent of the wave amplitude
$\epsilon$. Note, that for $\lambda\, r^2\to\infty$, $S\to 0$. This is 
in spite of the fact that spacetime approaches algebraic speciality. 
In fact, spacetime approaches algebraic speciality exponentially fast, 
as both $I$ and $J$ decay exponentially with $\lambda\, r^2$. This 
behavior {\em is} captured by the BB scalar $\xi$: from Eq.~(\ref{xi}), 
$\xi$ equals $I$ times a function of $S$ only. The speciality index $S$ is 
dropping off at great distances to zero like $\lambda^{-4}\, r^{-8}$, so that 
the function that multiplies $I$ approaches a constant. The BB scalar $\xi$ 
drops off exponentially, specifically like $\exp(-2\lambda\,r^2)$. 

\begin{figure}
\input epsf
\centerline{ \epsfxsize 10.0cm
\epsfbox{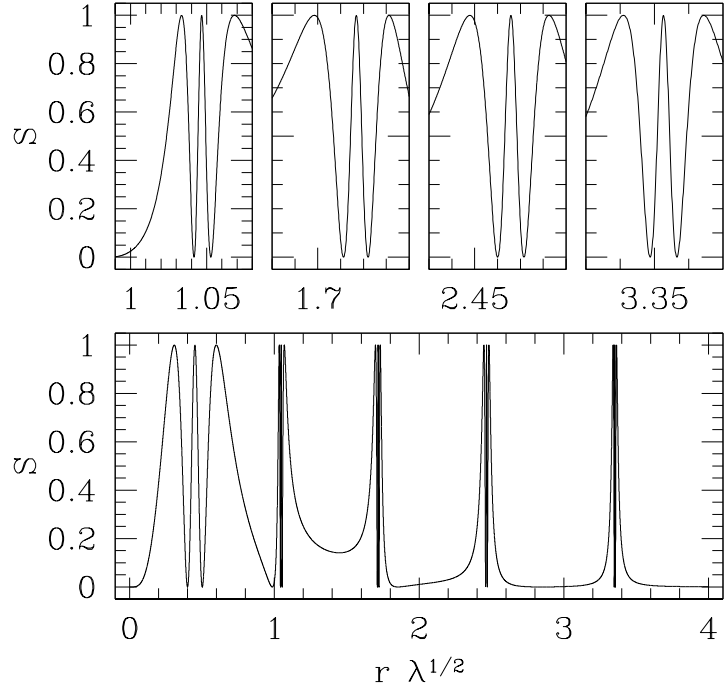}}
\caption{The speciality index $S$ as a function of $\lambda^{1/2}\, r$
on the equatorial plane, for linearized, even-parity, quadrupole
waves. The top panel magnifies the regions where $S$ oscillates
rapidly. }
\label{fig:quad-s}
\end{figure}

Figure \ref{fig:quad-s} shows the speciality index $S$ as a function
of $\lambda^{1/2}\, r$ on the equatorial plane.  There are 15 values
of $r$ for which $S=1$, arranged in five triplets. The values of $\rho
= \lambda^{1/2} r$ for which $S=1$ are
\begin{equation}
\begin{array}{cclll}
{_1}\rho_{1,2,3}&=&0.30800524515557994139, &0.45049403784384619791, 
&0.59829320761717622144\\
{_2}\rho_{1,2,3}&=&1.0335527869353531668, &1.0467779593567221747, 
&1.0682855799176748686\\
{_3}\rho_{1,2,3}&=&1.6984496413703925629, &1.7168422939913076313, 
&1.7311171061162415883\\
{_4}\rho_{1,2,3}&=&2.4478787994829345025, &2.4658595692782475639, 
&2.4824885271217855892\\
{_5}\rho_{1,2,3}&=&3.3409626177591728790, &3.3526507427990501828, 
&3.3643720160172523032\\
\end{array}
\end{equation}
In between points at which $S=1$, we have points at which 
$|S-1|=1$, which occur at the following locations:  
\begin{equation}
\begin{array}{cclll}
{_1}\rho^*_{1,2,3}&=&0, & 0.39945741403776425339, & 0.50160586730815486639 \\
{_2}\rho^*_{1,2,3}&=&0.98542263099668675629, & 1.0416944916187449946, & 
1.0527498193777135990 \\
{_3}\rho^*_{2,3}&=& & 1.7112566884721670706, & 1.7219786282054512104  \\
{_4}\rho^*_{1,2,3}&=&1.8534934086232694765, & 2.4600232831874878020, & 
2.4715457963797601620\\
{_5}\rho^*_{1,2,3}&=&2.8449085086360266876, & 3.3487504415432614190, & 
3.3565547217000555743 \\
\end{array}
\end{equation}
Notice that there is no value for ${_3}\rho^*_{1}$: the speciality index $S$ 
does not vanish between ${_2}\rho_{3}$ and ${_3}\rho_{1}$. In addition, 
$|S-1| \to 1$ also as $\lambda^{1/2}\, r\to\infty$. 

It appears that in some cases -- specifically for odd-parity
quadrupole waves -- the curvature invariant $J$ may vanish to
$O(\epsilon^3)$, so that its leading order is $O(\epsilon^4)$.
Generic gravitational waves include both polarization states, so that
the total curvature invariant $J$ is still of $O(\epsilon^3)$, and
$S=O(\epsilon^0)$.

\section{Results for Bowen--York Initial Data}
\label{app:B-Y}

Inserting (\ref{BY_K}) into
\begin{equation}
\B_{ij} = - \epsilon_{i}^{~~kl} \nabla_k K_{lj} 
 = - \gamma_{im} \epsilon^{mkl} (K_{lj,k} - K_{ln} \Gamma^n_{jk})
\end{equation}
and symmetrizing we find that the only non-vanishing coefficients of
the magnetic part of the Weyl tensor are
\begin{eqnarray}
B_{rr} & = & - \frac{64 \,\cos\theta}{(2\,r+M)^4} L \nonumber \\
&+& \frac{1536}{10}\,\frac{[16(2-3\,\cos^2\theta )r^3-20Mr^2-8M^2r-M^3]\,
\cos\theta}
{M\,(2\,r+M)^{10}}L^3 
\\
B_{r\theta} & = & - \frac{48\,r\,(2r-M)\,\sin\theta}{(2\,r+M)^5} \, L 
\nonumber \\
&+&\frac{384}{5}\, \frac{[40(1-3\,\cos^2\theta)r^4-20(7-3\,\cos^2\theta)Mr^3
-16M^2r^2+4M^3r+M^4]\,r\,\sin\theta}{(2\,r+M)^{11}}\, L^3
\\
B_{\theta\theta} & = & 48\, \frac{\,\cos\theta\,r^{2}}{(2\,r + M)^{4}}\, L 
\nonumber \\
&+& \frac {384}{5} \,\frac{\,\cos\theta \,[20\,M\,r^{2} + 8\,M^{2}\,r + M^{3} +
24\,r^{3}\,\cos^2\theta - 8\,r^{3}]\,r^{2}}{(2\,r + M)^{10}\,M}\, L^3
\\
B_{\phi \phi} & = & 
48\, \frac{\,\cos\theta\,\sin^2\theta\,r^{2}}{(2\,r + M)^{4}}\, L \nonumber \\
&+& \frac {384}{5} \,\frac{\,\cos\theta\,\sin^2\theta \,[
72\,r^{3}\,\cos^2\theta  - 56\,r^{3} + 20\,M\,r^{2} + 8\,M^{2}
\,r + M^{3}
]\,r^{2}}{(2\,r + M)^{10}\,M}\, L^3
\end{eqnarray}

The electric part of the Weyl tensor is computed from the 
Ricci tensor, whose non-vanishing components to $O(L^4)$ are
\begin{eqnarray}
R_{rr}&=& 
-\frac {8\,M}{(2\,r + M)^{2}\,r} \nonumber \\
&+& 
\frac {32}{5}\, \frac{ 144
\,r^{4}\,\cos^2\theta - 360\,r^{3}\,M\,\cos^2\theta + 
36\,r^{2}\,M^{2}\,\cos^2\theta - 32\,M^{2}\,r^{2} - 48\,r^{4}
 - 8\,M^{3}\,r + 440\,M\,r^{3} - M^{4}}{(2\,r + M)^{8}
\,M\,r}\, L^2 \nonumber \\
&+&
\frac{64}{25} \left[ - 8896\,M\,r^{6} + 16176\,M^{2}
\,r^{5} + 6480\,M^{3}\,r^{4} + 800\,M^{4}\,r^{3} - 36\,M^{5}\,r^{
2} - 12\,r\,M^{6} - M^{7} 
\mbox{} + 960\,r^{7} \right. \nonumber \\
&-& 10704\,r^{5}\,\cos^2\theta \,M
^{2} + 31872\,r^{6}\,\cos^2\theta \,M - 14400\,M\,r^{
6}\,\cos^4\theta  
\mbox{} + 2016\,M^{2}\,r^{5}\,\cos^4\theta  - 240\,M
^{4}\,r^{3}\,\cos^2\theta  \nonumber \\
&+& \left. 36\,M^{5}\,r^{2}\,
\cos^2\theta  - 3360\,r^{4}\,\cos^2\theta \,M^{3} 
\mbox{} - 5184\,r^{7}\,\cos^2\theta + 8064\,r^{7}\,
\cos^4\theta \right] \, L^{4} \left/ {\vrule
height0.44em width0em depth0.44em} \right. \!  \! [(2\,r + M)^{14
}\,M^{2}\,r] \nonumber \\
&+& 
\frac{4\,L^4}{2r+M}\,\left[\frac{2M(2r-M)}{r(2r+M)^2}
-\,\frac{\cot\theta}{r}\,\partial_{\theta}-2\,\frac{2r+3M}{2r+M}\,\partial_r
-\,\frac{1}{r}\,\partial_{\theta\theta}-2r\,\partial_{rr}\right]\,F(r,\theta)
\end{eqnarray}

\begin{eqnarray}
R_{r\theta}&=&
\frac {768}{5} \,\sin\theta\,\cos\theta \frac {r^{2}\,(2\,r - M)}{
(2\,r + M)^{7}\,M} \,L^2 \nonumber \\
&+& 
\frac {1536}{25} r^{2}\,\sin\theta\,
\cos\theta\,   
(156\,r^{4}\,\cos^2\theta - 52\,r^{4} - 78\,r^{3}\,
M\,\cos^2\theta + 186\,M\,r^{3} + 26\,M^{2}\,r^{2}
\nonumber \\ 
&-& 3\,M^{3}\,r - \,M^{4}) 
\, L^{4} \left/ {\vrule height0.44em width0em depth0.44em}
 \right. \!  \! [(2\,r + M)^{13}\,M^{2}]
+ \frac{4}{2r+M}\,L^4\,\left(\,2\frac{r-M}{2r+M}\,\partial_{\theta}-\,r
\,\partial_{r\theta}\,\right)\,F(r,\theta) 
\end{eqnarray}

\begin{eqnarray}
R_{\theta\theta} &=&
\frac {4\,M\,r}{(2\,r + M)^{2}} \nonumber \\
&-&
\frac {16}{5} \frac{(168\,r^{4}\,\cos^2\theta
- 72\,r^{4} - 304\,M\,r^{3} + 384\,r^{3}\,M\,\cos^2\theta + 
42\,r^{2}\,M^{2}\,\cos^2\theta - 38\,
M^{2}\,r^{2} 
- 8\,M^{3}\,r - M^{4})r}{(2\,r + M)^{8}\,M}\, L^2 \nonumber \\
&-&
\frac {32}{25} (1152r^7-M^7-3424Mr^6+1152M^2r^5+384M^3r^4-42M^5r^2-12M^6r
+8M^4r^3+13284r^7\,\cos^4\theta\nonumber \\
&-&11136\,r^7\,\cos^2\theta+8640\,r^6M\,\cos^4\theta
+3360\,r^6M\,\cos^2\theta+3456\,r^5M^2\,\cos^4\theta+2880\,r^5M^2\,\cos^2\theta
+2736\,r^4M^3\,\cos^2\theta\nonumber \\
&+&552\,r^3M^4\,\cos^2\theta+42\,r^2M^5\,\cos^2\theta)
r\, L^{4} \left/ {\vrule
height0.44em width0em depth0.44em} \right. \!  \! [M^{2}\,(2\,r
 + M)^{14}] \nonumber \\
&-&\frac{4r}{2r+M}\,L^4\,\left[\frac{M(2r-M)}{(2r+M)^2}+\,
\cot\theta\,\partial_{\theta}
+\frac{r(6r+M)}{2r+M}\,\partial_r+2\,\partial_{\theta\theta}+
r^2\,\partial_{rr}\,\right]
\,F(r,\theta) 
\end{eqnarray}

\begin{eqnarray}
R_{\phi\phi} &=&
\frac{4\,M\,r }{(2\,r + M)^{2}}\,\sin^2\theta \nonumber \\
&-&
\frac {16}{5} \,\sin^2\theta \frac{(336
\,r^{3}\,M\,\cos^2\theta + 30\,r^{2}\,M^{2}\,
\cos^2\theta + 120\,r^{4}\,\cos^2\theta 
 - 24\,r^{4} - 8\,M^{3}\,r 
- 256\,M\,r^{3} - 26\,M^{2}\,r^{2} - M^{4})\,r}{M\,(2
\,r + M)^{8}} \, L^2 \nonumber \\
&-&
\frac {32}{25} \,\sin^2\theta (
1440\,M^{2}\,r^{5}\,\cos^4\theta + 576\,M\,r^{6}\,
\cos^4\theta + 5760\,r^{7}\,\cos^4\theta + 408\,M^{4}\,r^{3}\,\cos^2\theta 
\nonumber \\
&-& 2688\,r^{
7}\,\cos^2\theta + 30\,M^{5}\,r^{2}\,\cos^2\theta + 
3648\,r^{5}\,\cos^2\theta \,M^{2} 
+ 2064\,r^{4}\,\cos^2\theta \,M^{3} + 10848\,
r^{6}\,\cos^2\theta \,M + 1056\,M^{3}\,r^{4} \nonumber \\
&+& 152\,M^{4}\,r^{3} - 2848\,M\,r^{6} + 2400\,M^{2}\,r^{5} + 768\,r^{7} - 30\,
M^{5}\,r^{2} - M^{7} - 12\,r\,M^{6})\,r \,L^{4} \left/ {\vrule
height0.44em width0em depth0.44em} \right. \!  \! [M^{2} (2\,r + M)^{14}]
\nonumber \\
&-&\frac{4r}{2r+M}\,\sin^2\theta\,L^4\,\left[\,\frac{M(2r-M)}{(2r+M)^2}+
\,r^2\,\partial_{rr}+2\,\cot\theta\,\partial_{\theta}+\,\frac{r(6r+M)}{2r+M}\,
\partial_r+
\,\partial_{\theta\theta}\,\right]\,F(r,\theta)\, ,
\end{eqnarray}
and the nonzero components of the electric part of the Weyl tensor are given by
\begin{eqnarray}
E_{rr} &=& 
-\frac{8M}{r\,(2r+M)^2}\nonumber \\
&+&
\frac {32}{5} \, \frac {144\,r^{4}
\,\cos^2\theta + 36\,r^{2}\,M^{2}\,\cos^2\theta 
- 48\,r^{4} - 32\,M^{2}\,r^{2} + 
80\,M\,r^{3} - 8\,M
^{3}\,r - M^{4}}{M\,r\,(2\,r + M)^{8}}\, L^2 \nonumber \\
&+&
\frac {64}{25} ( 704\,M\,r^{6} - 7824\,M^{2}
\,r^{5} - 3120\,M^{3}\,r^{4} - 400\,M^{4}\,r^{3} - 36\,M^{5}\,r^{2}
 - 12\,r\,M^{6} - M^{7} + 960\,r^{7} \nonumber \\
\mbox{} &+& 9792\,r^{7}\,\cos^4\theta - 6912\,r^{7}\,
\cos^2\theta - 8256\,r^{6}\,\cos^2\theta  
\,M + 16128\,M\,r^{6}\,\cos^4\theta  \nonumber \\
\mbox{} &+& 2448\,M^{2}\,r^{5}\,\cos^4\theta + 36\,M^{
5}\,r^{2}\,\cos^2\theta + 960\,M^{4}\,r^{3}\,
\cos^2\theta + 12864\,r^{5}\,\cos^2\theta 
\,M^{2} \nonumber \\
\mbox{} &+& 6240\,r^{4}\,\cos^2\theta \,M^{3})\, L^{4}
 \left/ {\vrule height0.44em width0em depth0.44em} \right. \!
 \! [M^{2}\,r\,(2\,r + M)^{14}]\nonumber \\
&+&\,\frac{4}{3}\,\frac{L^4}{2r+M}\,\left[\,\frac{6M(2r-M)}
{r(2r+M)^2}-2\,r\,\partial_{rr}+\frac{1}{r}\,\partial_{\theta\theta}+2\,
\frac{2r-5M}{2r+M}\,\partial_r+\,\frac{\cot\theta}{r}\,\partial_{\theta}\,
\right]
\,F(r,\theta) 
\end{eqnarray}

\begin{eqnarray}
E_{r\theta}&=&
\frac {768}{5} \,\sin\theta\,\cos\theta\,\frac {r^{2}
\,(2\,r - M)}{(2\,r + M)^{7}\,M}\, L^2 \nonumber \\
&+&
\frac {1536}{25}\, \sin\theta\,\cos\theta \, r^2 
(156\,r^{4}\,\cos^2\theta  - 52\,r^{4} - 78\,r^{3}\,
M\,\cos^2\theta  \nonumber \\
&+& 186\,M\,r^{3} + 26\,M^{2}\,r^{2}
 - 3\,M^{3}\,r - \,M^{4}) \,
L^{4} \left/ {\vrule height0.44em width0em depth0.44em}
 \right. \!  \! [(2\,r + M)^{13}\,M^{2}]\nonumber \\
&+&\,4\,\frac{L^4}{2r+M}\,\left(\,2\,\frac{r-M}{2r+M}\,\partial_{\theta}\,-\,
r\,\partial_{r\theta}\,\right)\,F(r,\theta)
\end{eqnarray}

\begin{eqnarray}
E_{\theta\theta}&=&
\frac{4Mr}{(2r+M)^2}\nonumber \\
&-&
\frac {16}{5}\,\frac{(168\,r^{4}\,\cos^2\theta - 72\,r^{4} - 304\,M\,r^{3} + 
384\,r^{3}\,M\,
\cos^2\theta  
+ 42\,r^{2}\,M^{2}\,\cos^2\theta - 38\,M^{2}
\,r^{2} - 8\,M^{3}\,r - M^{4})r}{M\,(2\,r + M)
^{8}}\, L^2 \nonumber \\
&-&
\frac {32}{25} \, ( 
23456\,M\,r^{6} - 66048\,M^{2}\,r^{5} - 26496\,M^{3}\,r^{4} - 3352\,
M^{4}\,r^{3} - 42\,M^{5}\,r^{2} - 12\,r\,M^{6} - M^{7} \nonumber \\
\mbox{} &+& 1152\,r^{7} + 13824\,r^{7}\,\cos^4\theta -
11136\,r^{7}\,\cos^2\theta + 29616\,r^{4}\,\cos^2\theta \,M^{3} \nonumber \\
\mbox{} &+& 89280\,M\,r^{6}\,\cos^4\theta + 3456\,M^{2
}\,r^{5}\,\cos^4\theta + 42\,M^{5}\,r^{2}\,\cos^2\theta  \nonumber \\
\mbox{} &-& 104160\,r^{6}\,\cos^2\theta \,M + 3912\,M^{4
}\,r^{3}\,\cos^2\theta + 70080\,r^{5}\,\cos^2\theta \,M^{2})r\,L^{4} \left/ 
{\vrule
height0.44em width0em depth0.44em} \right. \!  \! [M^{2} (2\,r + M)^{14}]
\nonumber \\
&+&\,\frac{4}{3}\,\frac{r}{2r+M}\,L^4\,\left[\,\cot\theta\,\partial_{\theta}
\,-\,\frac{r(2r-5M)}{2r+M}\,\partial_r-2\,\partial_{\theta\theta}
+\,r^2\,\partial_{rr}-3\,\frac{M(2r-M)}{(2r+M)^2}\,\right]\,F(r,\theta) 
\end{eqnarray}

\begin{eqnarray}
E_{\phi\phi}&=&
\frac{4Mr}{(2r+M)^2}\,\sin^2\theta \nonumber \\
&-&
\frac {16}{5} \,\sin^2\theta \,r( - 384\,r^{3}\,M\,\cos^2\theta + 
30\,r^{2}\,M^{2}\,\cos^2\theta \nonumber \\
\mbox{} &+& 120\,r^{4}\,\cos^2\theta - 26\,M^{2}\,r^{2
} - 24\,r^{4} - 8\,M^{3}\,r + 464\,M\,r^{3} - M^{4})\, L^{2} \left/
{\vrule height0.44em width0em depth0.44em} \right. \!  \! [M\,(2
\,r + M)^{8}] \nonumber \\
&-&
\frac {32}{25} \,\sin^2\theta \,r( - 57024\,M\,r^{6}\,\cos^4\theta + 
5760\,r^{7}\,\cos^4\theta + 1440\,M^{2}\,r^{5}\,\cos^4\theta  \nonumber \\
\mbox{} &-& 2688\,r^{7}\,\cos^2\theta - 17136\,r^{4}\,
\cos^2\theta \,M^{3} - 44352\,r^{5}\,\cos^2\theta \,M^{2} + 
30\,M^{5}\,r^{2}\,\cos^2\theta \nonumber  \\
\mbox{} &-& 1992\,M^{4}\,r^{3}\,\cos^2\theta + 87648\,r
^{6}\,\cos^2\theta \,M + 20256\,M^{3}\,r^{4} - 22048\,M
\,r^{6} + 768\,r^{7} \nonumber \\
\mbox{} &-& 30\,M^{5}\,r^{2} - M^{7} + 2552\,M^{4}\,r^{3} + 50400\,M
^{2}\,r^{5} - 12\,r\,M^{6})\, L^{4} \left/ {\vrule
height0.44em width0em depth0.44em} \right. \!  \! [M^{2}\,(2\,r
 + M)^{14}]\nonumber \\
&-&\,\frac{4}{3}\,\frac{r}{2r+M}\,\sin^2\theta\,L^4\,\left[\frac{3M(2r-M)}
{(2r+M)^2}-\,r^2\,\partial_{rr}-\,\partial_{\theta\theta}+\,\frac{r(2r-5M)}
{2r+M}\,\partial_r+2\,\cot\theta\,\partial_{\theta}\,\right]\,F(r,\theta) \, .
\end{eqnarray}

Next, we find the tensor $C_{ab}$ to $O(L^4)$, from which we construct the 
Weyl tensor invariants $I$ and $J$. To $O(L^4)$, we have
\begin{eqnarray}
I &=& 
\frac {12288\,M^{2}\,r^{6}}{(2\,r + M)^{12}}
+\, 294912\,i\,\frac{\,\cos\theta Mr^7}{(2r+M)^{14}}\, L\nonumber \\
&-&
\frac {294912}{5}\, \frac{(128\,r^{4}\,\cos^2\theta+ 24\,r^{4} + 
144\,r^{3}\,M\,\cos^2\theta 
- 8\,M\,r^{3} + 32\,r^{2}\,M^{2}\,\cos^2\theta 
- 14\,M^{2}\,r^{2} - 8\,M^{3}\,r - M^{4})r^{6}}{(2\,r + M)^{18}} \, L^2 
\nonumber \\
&-&
\frac{1179648}{5}\,i\,\,\cos\theta\,
(80\,r^{4}\,\cos^2\theta - 128\,r^{3}
\,M\,\cos^2\theta  
+ 20\,r^{2}\,M^{2}\,\cos^2\theta + 16\,r^{4}
 + 112\,M\,r^{3} - 136\,M^{2}\,r^{2} - 7\,M^{4} \nonumber \\
&-& 56\,M^{3}\,r)\,r^7\, L^3 
\left/ {\vrule height0.44em width0em depth0.44em}
 \right. \!  \! [(2\,r + M)^{20}\,M]\nonumber \\
&+&
{\displaystyle \frac {196608}{25}} (300\,M^{7}\,r + 19\,M^{8} -
32288\,M^{2}\,r^{6} + 29712\,M^{3}\,r^{5} + 21312\,M\,r^{7} + 12972
\,M^{4}\,r^{4}\nonumber \\
\mbox{} &+& 1548\,M^{6}\,r^{2} + 4672\,M^{5}\,r^{3} + 2496\,r^{8}
 - 1920\,r^{8}\,\cos^2\theta + 8640\,r^{8}\,\cos^4\theta \nonumber \\
\mbox{} &+& 43776\,r^{7}\,M\,\cos^2\theta - 98328\,r^{
4}\,M^{4}\,\cos^2\theta - 158016\,r^{7}\,M\,\cos^4\theta  \nonumber\\
\mbox{} &-& 39504\,r^{5}\,M^{3}\,\cos^4\theta - 1356\,
r^{2}\,M^{6}\,\cos^2\theta + 540\,r^{4}\,M^{4}\,
\cos^4\theta  \nonumber\\
\mbox{} &-& 196608\,r^{5}\,M^{3}\,\cos^2\theta - 33408
\,r^{6}\,M^{2}\,\cos^2\theta - 233184\,r^{6}\,M^{2}
\,\cos^4\theta  \nonumber\\
\mbox{} &-& 19056\,r^{3}\,M^{5}\,\cos^2\theta )\,r^{6}\,L
^{4} \left/ {\vrule height0.44em width0em depth0.44em}
 \right. \!  \! [(2\,r + M)^{24}\,M^{2}]\,\nonumber \\
&-&\,4096\,\frac{Mr^6}{(2r+M)^{11}}\,L^4\,\left[\,\cot\theta\,\partial_{\theta}
+2\,\frac{r(2r-5M)}{2r+M}\,\partial_r+\,\partial_{\theta\theta}-\,2\,r^2\,
\partial_{rr}
+6\,\frac{M(10r-M)}{(2r+M)^2}\,\right]\,F(r,\theta) 
\label{I_expansion}
\end{eqnarray}
and
\begin{eqnarray} 
J &=& 
\frac {262144\,M^{3}\,r^{9}}{(2\,r + M)^{18}}\,+\,9437184\,i\,
\cos\theta\,\frac{M^2r^{10}}{(2r+M)^{20}}\, L\nonumber \\
&-&
\frac {9437184}{5} \,\frac{(248\,r^{4}\,\cos^2\theta + 24\,r^{4} + 
264\,r^{3}\,M\,\cos^2\theta 
- 8\,M\,r^{3} + 62\,r^{2}\,M^{2}\,\cos^2\theta 
- 14\,M^{2}\,r^{2} - 8\,M^{3}\,r - M^{4})M\,r^{9}}
{(2\,r + M)^{24} }\, L^2\nonumber \\
&-&
\frac{75497472}{5}\,i\,\cos\theta\,
(172\,r^{4}\,\cos^2\theta + 43
\,r^{2}\,M^{2}\,\cos^2\theta + 92\,r^{3}\,M\,
\cos^2\theta - 40\,M^{3}\,r 
+ 44\,r^{4} + 44\,M\,r^{3} - 5\,M^{4} \nonumber \\
&-& 89\,M^{2}\,r^{2})r^{10}\, L^3
\left/ {\vrule height0.44em width0em depth0.44em}
 \right. \!  \! (2\,r + M)^{26} \nonumber \\
&+&
{\displaystyle \frac {25165824}{25}} (111\,M^{7}\,r + 7\,M^{8}
 - 74096\,M^{2}\,r^{6} + 13116\,M^{3}\,r^{5} + 28656\,M\,r^{7}
 + 3720\,M^{4}\,r^{4} \nonumber\\
\mbox{} &+& 594\,M^{6}\,r^{2} + 1708\,M^{5}\,r^{3} + 10752\,r^{8}
\,\cos^2\theta - 31728\,r^{7}\,M\,\cos^4\theta  \nonumber\\
\mbox{} &-& 153264\,r^{6}\,M^{2}\,\cos^4\theta - 7932
\,r^{5}\,M^{3}\,\cos^4\theta + 1800\,r^{4}\,M^{4}\,
\cos^4\theta  \nonumber\\
\mbox{} &-& 12504\,r^{3}\,M^{5}\,\cos^2\theta - 978\,
r^{2}\,M^{6}\,\cos^2\theta - 60240\,r^{4}\,M^{4}\,
\cos^2\theta  \nonumber\\
\mbox{} &+& 10944\,r^{7}\,M\,\cos^2\theta + 129312\,r
^{6}\,M^{2}\,\cos^2\theta - 122160\,r^{5}\,M^{3}\,
\cos^2\theta - 384\,r^{8} \nonumber\\
\mbox{} &+& 28800\,r^{8}\,\cos^4\theta )r^{9}\,L^{4}
 \left/ {\vrule height0.44em width0em depth0.44em} \right. \!
 \! [(2\,r + M)^{30}\,M]\nonumber \\
&-&\,131072\,\frac{M^2r^9}{(2r+M)^{17}}\,L^4\,
\left[\,\cot\theta\,\partial_{\theta}
+2\,\frac{r(2r-5M)}{2r+M}\,\partial_r+\,\partial_{\theta\theta}-\,
2\,r^2\,\partial_{rr}
+6\,\frac{M(10r-M)}{(2r+M)^2}\,\right]\,F(r,\theta)\, .
\label{J_expansion}
\end{eqnarray}

\section{The Bowen--York BB radiation scalar is independent 
of $F(r,\theta)$}
\label{app:BY_proof}

In this Appendix we outline a calculation that shows why the
speciality index $S$, and hence the BB radiation scalar $\xi^0$, is
independent of the $O(L^4)$ terms in the conformal factor.

Let us expand the curvature invariants $I$ and $J$ to $O(L^4)$ as
\begin{equation}
I=\sum_{k=0}^{4}i_k\,L^k \;\;\; , \;\;\; J=\sum_{k=0}^{4}j_k\,L^k\, ,
\end{equation}
where the expansion coefficients $i_k$ and $j_k$ are given by
(\ref{I_expansion}) and (\ref{J_expansion}).  
The only coefficients that depend on $F(r,\theta)$ are $i_4$ and
$j_4$.  The speciality index $S$ is given by (\ref{S}) and can also be
expanded in powers of $L$ up to order $O(L^4)$ as
\begin{equation}
S=\sum_{k=0}^{4}s_k\,L^k\,.
\end{equation}
The expansion coefficients can be found from the definition
of $S$ and the expansion coefficients for $I$ and $J$.

The leading order term in $S$ is unity.  A straightforward
calculation shows that the next contributing order is at
$O(L^4)$.  What remains to be shown is that the $O(L^4)$ 
term $s_4 L^4$ is independent of
$F(r,\theta)$.  It turns out that the only dependence of $s_4$
on either $i_4$ or $j_4$ is through the combination
$2\,i_0\,j_4\,-\,3\,j_0\,i_4.$  A direct substitution finds that
this combination equals exactly zero, which completes the proof.

The BB scalar $\xi^0$ is proportional to $I$ times a power series in
$S-1$, the leading order term being of $O(S-1)$. Consequently, the
earliest order at which $F(r,\theta)$ can contribute is at
$O(L^8)$.  Specifically, it does not contribute at $O(L^4)$, which is
the leading order for the BB scalar.

\end{appendix}

\end{document}